\makeatletter
\def\input@path{{template/}}
\makeatother
\documentclass{ieeeaccess}
\usepackage{cite}
\usepackage{amsmath,amssymb,amsthm,mathtools}
\usepackage{graphicx}
\graphicspath{{fig/}{template/}}
\usepackage{afterpage}
\usepackage{textcomp}
\usepackage{hyperref}
\hypersetup{hidelinks,hypertexnames=false}
\usepackage[english]{babel}
\usepackage{bm}
\usepackage{multicol}
\usepackage{cuted}
\usepackage{algorithm,algpseudocode}
\usepackage[acronym,shortcuts]{glossaries}
\glsdisablehyper
\usepackage{tabularx}
\usepackage{soul}
\usepackage{subfigure}

\makeatletter
\patchcmd{\thebibliography}
  {\section*{\refname}}
  {\section*{\refname}\Hy@raisedlink{\hyper@anchorstart{\@currentHref}\hyper@anchorend}}
  {}{}
\expandafter\patchcmd\csname\string\IEEEbiography\endcsname
  {\vskip \@IEEEBIOskipN plus 1fil minus 0\baselineskip}
  {\vskip \@IEEEBIOskipN}
  {}{}
\makeatother

\newcommand{\orcidlogo}{%
  \leavevmode\hbox to 7.2bp{%
    \vrule width0pt height7.2bp depth0pt%
    \pdfliteral{q
      .028125 0 0 -.028125 0 7.2 cm
      .65098 .80784 .22353 rg
      256 128 m 256 198.7 198.7 256 128 256 c
      57.3 256 0 198.7 0 128 c
      0 57.3 57.3 0 128 0 c
      198.7 0 256 57.3 256 128 c h f
      1 1 1 rg
      70.9 79.1 m 86.3 79.1 l 86.3 186.2 l 70.9 186.2 l h
      108.9 79.1 m 150.5 79.1 l
      190.1 79.1 207.5 107.4 207.5 132.7 c
      207.5 160.2 186 186.3 150.7 186.3 c
      108.9 186.3 l h
      124.3 172.4 m 148.8 172.4 l
      183.7 172.4 191.7 145.9 191.7 132.7 c
      191.7 111.2 178 93 148 93 c
      124.3 93 l h f
      88.7 56.8 m 88.7 62.3 84.2 66.9 78.6 66.9 c
      73 66.9 68.5 62.3 68.5 56.8 c
      68.5 51.2 73 46.7 78.6 46.7 c
      84.2 46.7 88.7 51.3 88.7 56.8 c h f
      Q}%
    \hss}}
\newcommand{\orcidid}[1]{\kern.08em\href{https://orcid.org/#1}{\raisebox{.55ex}{\orcidlogo}}}

\newacronym{MU}{MU}{multi-user}
\newacronym{PLS}{PLS}{physical layer security}
\newacronym{adc}{ADC}{analogue/digital converter}
\newacronym{cpu}{CPU}{central processing unit}
\newacronym{dac}{DAC}{digital/analogue converter}
\newacronym{dsp}{DSP}{digital signal processor}
\newacronym{dsrc}{DSRC}{dedicated short-range communication}
\newacronym{fpga}{FPGA}{field-programmable gate array}
\newacronym{gfsk}{GFSK}{Gaussian frequency shift keying}
\newacronym{gmsk}{GMSK}{Gaussian minimum shift keying}
\newacronym{gsm}{GSM}{global system for mobile communications}
\newacronym{gpp}{GPP}{general purpose processor}
\newacronym{gui}{GUI}{graphical user interface}
\newacronym{ieee}{IEEE}{institute of electrical and electronics engineers}
\newacronym{io}{I/O}{input/output}
\newacronym{iot}{IoT}{internet of things}
\newacronym{DCA}{DCA}{difference-of-convex algorithmic}
\newacronym{nfc}{NFC}{near-field communications}
\newacronym{los}{LOS}{line-of-sight}
\newacronym{AoD}{AoD}{angle of departure}
\newacronym{AoA}{AoA}{angle of arrival}
\newacronym{pll}{PLL}{phase-locked loop}
\newacronym{qam}{QAM}{quadrature amplitude modulation}
\newacronym{rf}{RF}{radio frequency}
\newacronym{sdr}{SDR}{software-defined radio}
\newacronym{SKG}{SKG}{secret key generation}
\newacronym{snr}{SNR}{signal-to-noise ratio}
\newacronym{soc}{SoC}{system on chip}
\newacronym{src}{SRC}{short-range communication}
\newacronym{usb}{USB}{universal serial bus}
\newacronym{usrp}{USRP}{universal software radio peripheral}
\newacronym{NG-IoT}{NG-IoT}{next-generation internet of things}
\newacronym{5G}{5G}{fifth-generation}
\newacronym{6G}{6G}{sixth-generation}
\newacronym{B5G}{B5G}{beyond 5G}
\newacronym{MMSE}{MMSE}{minimum mean square error}
\newacronym{SRM}{SRM}{sum-rate maximization}
\newacronym{SecLM}{SecLM}{secrecy-enhancement via leakage minimization}
\newacronym{LMM}{LMM}{lagrangian multiplier method}
\newacronym{SGD}{SGD}{stochastic gradient descent}
\newacronym{BS}{BS}{bisection}
\newacronym{AP}{AP}{access point}
\newacronym{DL}{DL}{downlink}
\newacronym{UL}{UL}{uplink}
\newacronym{SecBF}{SecBF}{secrecy-maximization beamforming}
\newacronym{BF}{BF}{beamforming}
\newacronym{TX}{TX}{transmit}
\newacronym{RX}{RX}{receive}
\newacronym{CSI}{CSI}{channel state information}
\newacronym{DoF}{DoF}{degrees-of-freedom}
\newacronym{CF-mMIMO}{CF-mMIMO}{cell-free massive multiple-input multiple-output}
\newacronym{MIMO}{MIMO}{multiple-input multiple-output}
\newacronym{MISO}{MISO}{multiple-input single-output}
\newacronym{SISO}{SISO}{single-input single-output}
\newacronym{mu-MIMO}{mu-MIMO}{multi-user MIMO}
\newacronym{mu-MISO}{mu-MISO}{multi-user MISO}
\newacronym{AWGN}{AWGN}{additive white Gaussian noise}
\newacronym{SIC}{SIC}{self-interference cancellation}
\newacronym{SINR}{SINR}{signal to interference plus noise ratio}
\newacronym{SNR}{SNR}{signal to noise ratio}
\newacronym{PSD}{PSD}{positive semidefinite}
\newacronym{SDP}{SDP}{semidefinite programming}
\newacronym{MRC}{MRC}{maximum ratio combining}
\newacronym{SotA}{SotA}{state-of-the-art}
\newacronym{D2D}{D2D}{device-to-device}
\newacronym{PhySec}{PhySec}{physical layer security}
\newacronym{RIS}{RIS}{reflective intelligent surface}
\newacronym{BDRIS}{BD-RIS}{beyond-diagonal reconfigurable intelligent surface}
\newacronym{P2P}{P2P}{point-to-point}
\newacronym{QoS}{QoS}{quality-of-service}
\newacronym{CCP}{CCP}{concave-convex procedure}
\newacronym{NN}{NN}{neural network}
\newacronym{TDD}{TDD}{time division duplex}
\newacronym{NMSE}{NMSE}{normalized mean squared error}
\newacronym{FP}{FP}{fractional programming}
\newacronym{QT}{QT}{quadratic transform}
\newacronym{LDT}{LDT}{Lagrangian dual transform}
\newacronym{BER}{BER}{bit error rate}
\newacronym{SER}{SER}{symbol error rate}
\newacronym{CDF}{CDF}{cumulative distribution function}
\newacronym{CDM}{CDM}{coordinate descent method}
\newacronym{CGA}{CGA}{conjugate gradient ascent}
\newacronym{UE}{UE}{user equipment}

\newcommand{\bHeq}[1]{{\widetilde{\mathbf{H}}_{#1}}}
\newcommand{\bHeqAP}[2]{{\widetilde{\mathbf{H}}_{#1,#2}}}
\newcommand{\bHTX}[1]{{\mathbf{H}_{\mathrm{TX},#1}}}
\newcommand{\bHRX}[1]{{\mathbf{H}_{\mathrm{RX},#1}}}
\newcommand{\bTheta}{{\bm{\Theta}}}

\renewcommand*{\IEEEeqnarraydecl}{%
\setlength{\jot}{2\IEEEnormaljot}
}

\def\BibTeX{{\rm B\kern-.05em{\sc i\kern-.025em b}\kern-.08em
    T\kern-.1667em\lower.7ex\hbox{E}\kern-.125emX}}
\begin{document}
\history{Date of publication xxxx 00, 0000, date of current version xxxx 00, 0000.}
\doi{10.1109/ACCESS.2026.XXXXXXX}

\markboth
{I. A. M. Sandoval \emph{et al.}: Low-Complexity Leakage Minimization Beamforming}
{I. A. M. Sandoval \emph{et al.}: Low-Complexity Leakage Minimization Beamforming}

\title{Low-complexity Leakage Minimization Beamforming for Large-scale Multi-user Cell-Free Massive MIMO}

\author{Iv{\'a}n Alexander Morales Sandoval\orcidid{0000-0002-8601-5451}\authorrefmark{1}, \IEEEmembership{Graduate Student Member, IEEE}, \\
{Marko Fidanovski\orcidid{0009-0005-2926-1604}\authorrefmark{1}, \IEEEmembership{Graduate Student Member, IEEE}}, \\
Getuar Rexhepi\orcidid{0009-0002-3268-522X}\authorrefmark{1}, \IEEEmembership{Graduate Student Member, IEEE},
Kengo Ando\orcidid{0000-0003-0905-2109}\authorrefmark{1}, \IEEEmembership{Member, IEEE}, and \\
Giuseppe Thadeu Freitas de Abreu\orcidid{0000-0002-5018-8174}\authorrefmark{1}, \IEEEmembership{Senior Member, IEEE}}
\address[1]{School of Computer Science and Engineering, Constructor University, 28759 Bremen, Germany}
\corresp{Corresponding author: Giuseppe Thadeu Freitas de Abreu (email: gabreu@constructor.university).}

\begin{abstract}
We propose a low-complexity \ac{BF} scheme for secrecy-rate maximization in \ac{MU} \ac{CF-mMIMO} systems, where legitimate users may act as non-colluding eavesdroppers of one another.
To this end, we formulate an information-leakage minimization problem and cast it into a tractable \ac{DCA} form by leveraging \ac{FP}.
The resulting non-convex problem is solved through a \ac{CCP}-based beamformer update{, and an additional row-wise \ac{CDM} implementation is introduced to avoid explicit matrix inversion in the dominant linear-solve step}.
{Additionally, we consider both direct \ac{TX}-\ac{BF} and \ac{BDRIS}-assisted operation by defining an equivalent channel between each access point and user that combines the direct and \ac{RIS}-assisted propagation components.}
Simulation results show that the proposed \ac{SecLM}-\ac{BF} framework achieves secrecy and sum-rate performance close to \ac{SotA} \ac{SDP}-based benchmarks{, as well as \ac{FP}-based benchmarks, while providing a scalable inversion-free implementation for large-scale secure \ac{CF-mMIMO} deployments}.
\end{abstract}

\begin{keywords}
Physical layer security, beamforming design, convex optimization, cell-free massive MIMO.
\end{keywords}

\titlepgskip=-21pt
\maketitle
\glsresetall

\section{Introduction}
\label{sec:introduction}

The exponential growth in wireless connectivity over the past decade has resulted in transformative advances in communication technologies.
With the transition from \ac{5G} to \ac{6G} wireless systems already underway, forecasts predict that over five billion devices will be wirelessly connected in the coming years \cite{Wang2023}.
Within such a paradigm, new requirements and challenges emerge, which are driven not only by the number of devices but also by their diverse capabilities and the complex \ac{QoS} demands they impose.
In particular, future wireless networks must deliver high reliability, low latency, and support a wide range of connected devices.

In response to these demands, \ac{CF-mMIMO} systems have emerged as a promising architectural paradigm.
Unlike traditional cellular structures, \ac{CF-mMIMO} systems are characterized by a large number of geographically distributed \acp{AP} that jointly serve users without the \mbox{constraints of predefined} cell boundaries.
This architecture enhances spectral efficiency, spatial diversity, and user fairness by harnessing the full spatial \acp{DoF} offered by the network \cite{Ngo15}.

However, the benefits of \ac{CF-mMIMO} come at the expense of increased system complexity and novel security threats.
Specifically, the broadcast nature of the wireless medium—coupled with the overlapping coverage provided by distributed \acp{AP}—creates significant challenges in ensuring secure communication, particularly in \ac{MU} environments\cite{Douong2019}.
One of the critical threats in \ac{CF-mMIMO} systems is \textit{information leakage}, which refers to the unintended reception of transmitted signals by users to whom the information is not intended.
This phenomenon is especially concerning in the \ac{DL} scenario, where each user may act as a passive, yet opportunistic, eavesdropper\cite{Park2024,Timilsina2018}.

Unlike conventional models that assume an external, malicious eavesdropper with a known channel, the legitimate-yet-untrusted user model better reflects the realities of dense network deployments, where users may seek to gain access to information directed to others, without actively disrupting the system.
In such a setting, the design of \ac{TX} \ac{BF} strategies must account for both maximizing intended signal power and minimizing signal leakage to unintended users, thereby ensuring information-theoretic privacy guarantees at the physical layer\cite{khisti10-I,khisti10-II,shenHong19,li2010}.

\Ac{PhySec}, which aims to enhance security by exploiting inherent characteristics of the wireless channel, has been widely explored as a complementary mechanism to traditional cryptographic techniques.
Among others \ac{PhySec} techniques, \ac{SecBF} has proved particularly promising.
The central idea behind \ac{SecBF} is to tailor the spatial transmission patterns such that the intended user's channel is enhanced while signal reception at potential eavesdroppers is suppressed.
Existing \ac{SecBF} methods, however, often suffer from critical limitations: they either rely on computationally intensive convex optimization methods such as \ac{SDP} \cite{Li2018}, assume simplified system models, $e.g.$ point-to-point \ac{MISO} or \ac{SISO} configurations), or operate under the unrealistic assumption of known eavesdropper channels.

Furthermore, most current designs impose per-user power constraints or reduce system complexity by employing beam domain compression, both of which compromise the optimality and scalability of the solution.
These constraints are particularly detrimental in \ac{CF-mMIMO} systems, where the flexibility and coordination between \acp{AP} are central to achieving performance gains. 
A viable solution for real-world systems must therefore balance the conflicting goals of communication performance, secrecy enhancement, and algorithmic efficiency.

{
\Acp{RIS} provide an additional means of improving this balance by shaping the propagation environment rather than relying only on active transmit-side processing.
In particular, \acp{BDRIS} generalize conventional diagonal \ac{RIS} architectures by allowing controlled coupling among surface elements through a non-diagonal scattering matrix, which offers additional spatial degrees of freedom for improving robustness, achievable rate, and energy efficiency in challenging propagation conditions \cite{FidanovskiArx2026}.
The reciprocal \ac{BDRIS} design in \cite{FidanovskiArx2026} focuses on physically consistent scattering-matrix optimization via manifold methods, whereas the related study in \cite{SandovalArx2026} considers a \ac{BDRIS}-aided \ac{MU} \ac{CF-mMIMO} communication setting close to the one considered here and investigates the corresponding link performance through distributed scattering-matrix design and \ac{MIMO} \ac{BF}.
These observations motivate evaluating the proposed leakage-minimization beamformer both for direct channels between the \acp{AP} and users and for equivalent channels that include a \ac{BDRIS}-assisted reflected path.
}

Recently, \ac{FP} has emerged as a powerful tool for solving non-convex optimization problems, particularly in the context of wireless communications.
Proposed in \cite{shen2018-I}, \ac{FP} allows for the reformulation of non-convex objectives, such as the ratio of signal-to-interference-plus-noise, into tractable surrogate problems that can be efficiently solved.
Particularly in \ac{BF} design, \ac{FP} has been shown to yield significant performance improvements while maintaining low computational complexity, making it an attractive alternative to traditional methods\cite{shen2018-I,shen2018-II}.
On the other hand, difference of convex problems is yet another class of non-convex optimization problems that have been studied in the past.
The \ac{CCP} is a well-known technique for solving such problems, which iteratively refines the solution by approximating the non-convex objective with a convex surrogate \cite{yuille03}.
As it can be seen in the literature, \ac{CCP} has been successfully applied to various wireless communication problems, including wireless localization \cite{tomic2025} and security in \ac{iot} networks \cite{haider2019}.

In the light of above, this paper aims to utilize these insights to address key limitations of existing methods by proposing a novel \textit{low-complexity beamforming framework based on \ac{FP} and \ac{CCP}} for information leakage minimization in CF-mMIMO systems. 
The proposed approach builds upon recent advances in FP-based optimization, which allow for non-convex objectives—such as the ratio of signal-to-interference-plus-noise—to be reformulated into tractable surrogate problems. 

Unlike traditional secrecy beamforming designs that rely on high-complexity SDP solvers, our method exploits the structure of the fractional objective and introduces a power-optimized and leakage-aware FP-based beamformer that significantly reduces computational overhead while maintaining strong secrecy performance.
The performance of the proposed \ac{BF} algorithm is verified through numerical simulations, under the realistic assumption of spatial channel correlation and presence of opportunistic eavesdroppers.
As a result of combining \ac{FP} and \ac{CCP} methods, the proposed algorithm designes \ac{BF} matrices that achieve similar secrecy rate to the \ac{SotA} algorithms, thus providing a trade-off of performance and computational complexity.

The rest of the paper is organized as follows. Section~II describes the system model and problem formulation. 
Section~III reviews related work on SecBF and FP-based optimization.
In Section~IV, we present our proposed FP-based leakage minimization beamforming framework and provide a complexity analysis for the proposed method. Section~V includes simulation results that validate the effectiveness of our method.
Section~VI concludes the paper with final remarks and potential directions for future research.

\noindent \textit{\textbf{Notation: }}
Column vectors and matrices are respectively denoted by lower- and upper-case bold face letters.
The $\ell_2$ and Frobenius norms are denoted by $\|\cdot\|_2$ and $\|\cdot\|_\mathrm{F}$, respectively.
The transpose operation is indicated by the superscript $^\mathrm{T}$.
The real part of a complex scalar is denoted as $\Re\{\cdot\}$, the conjugate transpose operation is indicated by the superscript $^\mathrm{H}$ and the inverse of the conjugate transpose by $^\mathrm{-H}$.
The circularly symmetric complex Gaussian distribution with mean $\nu$ and variance $\sigma^2$ is denoted by $\mathcal{CN}(\nu,\sigma^2)$.
The log determinant function is denoted by $\log_n|\cdot|$.
The non-negative operator is denoted by $(\cdot)^+$ and is equivalent to $\max(\cdot,0)$.
Subscripts in lower-case math (italicized) font denote indices, while subscripts in upper-case and occasionally in lower-case text (upright) font are used to contextualize variables. Superscripts enclosed in parenthesis indicate iteration indices.

\vspace{-2ex}
\section{Preliminaries}
\label{sec:preliminaries}
Before presenting the system model and problem statement, it is important to clarify the context and scope of this work.
Specifically, we note that uplink beamforming (\ac{UL}-\ac{BF}) for leakage minimization is not considered here, as it would require the unrealistic assumption that each $k$-th \ac{UE} possesses knowledge of its channels to all other \acp{UE}.

\vspace{-2ex}
\subsection{System Model}
\label{subsec:sysmodel}

\begin{figure}[t]
\centering
\includegraphics[width=0.9\columnwidth]{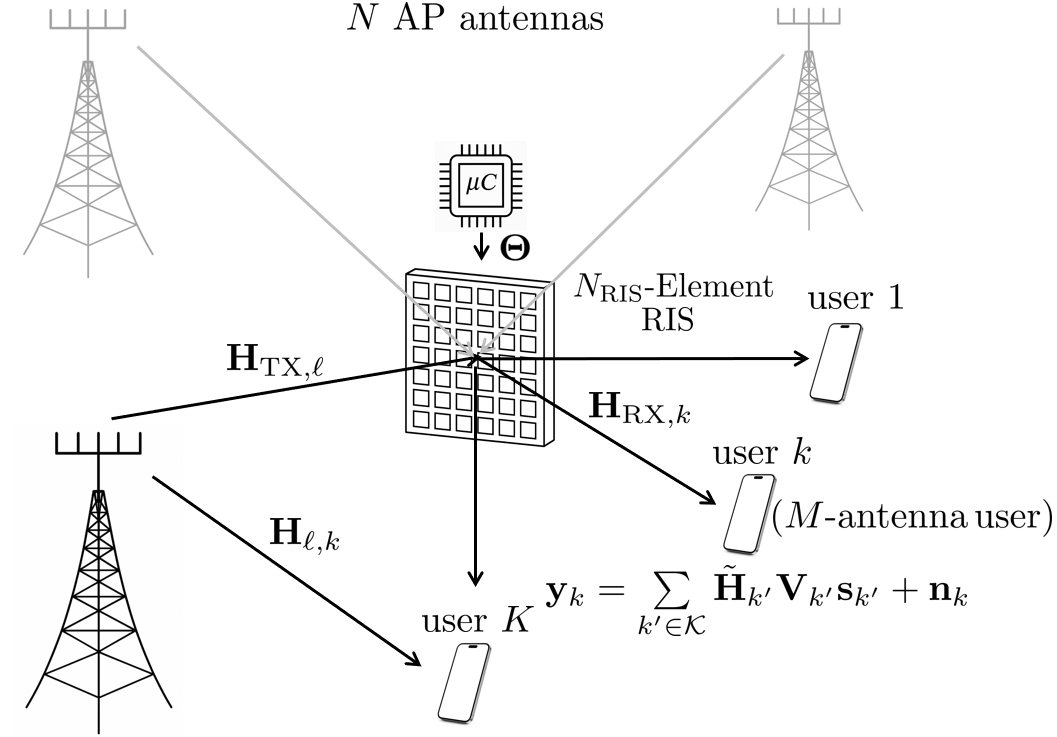}
\caption{An \ac{MU} \ac{CF-mMIMO} \ac{DL} system where a set of $L$ \acp{AP} equipped with $N$ antennas cooperatively serve $K$ \acp{UE} with $M$ antennas each.}
\label{fig:system_model}
\vspace{-2ex}
\end{figure}

We consider a downlink (DL) cell-free massive multiple-input multiple-output (CF-mMIMO) system, as depicted in Fig.~\ref{fig:system_model}, comprising $N$ transmit antennas distributed across $L$ access points (APs), each equipped with $N_t$ transmit antennas (i.e., $N = L \times N_t$). 
The system serves $K$ user equipments (UEs), each equipped with $M$ receive antennas. Communication from the collective APs to each UE is intended to be confidential, treating the remaining $K-1$ UEs as passive, non-colluding eavesdroppers who may potentially intercept unintended information.

Let $\mathcal{L} \triangleq \{1, \ldots, L\}$ and $\mathcal{K} \triangleq \{1, \ldots, K\}$ denote the index sets for the $L$ APs and $K$ UEs, respectively. The channel between the $\ell$-th AP and the $k$-th UE is represented by $\mathbf{H}_{\ell,k} \in \mathbb{C}^{M \times N_t}$. When spatial correlation is present, $\mathbf{H}_{\ell,k}$ can be modeled as in~\cite{bolcskei03, xiong22}:
\begin{equation}
\label{eq:chan_model}
\mathbf{H}_{\ell,k} \triangleq  \sqrt{\,\mathbf{R}_{\ell,k}} \mathbf{G}_{\ell,k} \sqrt{\,\mathbf{T}_{\ell,k}}^{\mathrm{T}},
\end{equation}
where $\mathbf{G}_{\ell,k} \in \mathbb{C}^{M \times N_t} \sim \mathcal{CN}(0,\sigma^2\bm{I})$ represents the small-scale fading effects, while the matrices $\mathbf{R}_{\ell,k} \in \mathbb{C}^{M \times M}$ and $\mathbf{T}_{\ell,k} \in \mathbb{C}^{N_t \times N_t}$ account for the spatial correlation at the receiver and transmitter sides of the channel between the $\ell$-th \ac{AP} and the $k$-th \ac{UE}, respectively.

The spatial correlation matrices $\mathbf{R}_{\ell,k}$ and $\mathbf{T}_{\ell,k}$ in equation \eqref{eq:chan_model} are determined via the local scattering model described in \cite{SIG-093}, which can be briefly summarized as follows.
Let the \ac{AoD} and \ac{AoA} of a path between the $\ell$-th \ac{AP} and the $k$-th \ac{UE} be respectively denoted by $\varphi^{\mathrm{Tx}}_{\ell,k}$ and $\varphi^{\mathrm{Rx}}_{\ell,k}$.
Briefly omitting the super- and sub-scripts for conciseness, each of the angles $\varphi$ are assumed to be Gaussian random variables with mean {$\mu_\phi$} and standard deviation {$\sigma_\phi$}, with the pair {$(\mu_\phi,\sigma_\phi)$} associated to a given scattering cluster, such that the distribution of $\varphi$ is given by
\begin{equation}
f (\varphi) = \tfrac{1}{\sqrt{2 \sigma_\phi}} e ^ {-\frac{(\phi-\mu_\phi)^2}{2\sigma_\phi^2}}.
\end{equation} 

Then, using the operator $[ \mathbf{X} ]_{q,m}$ to denote the $q$-th row and $m$-th column elements of some matrix $\mathbf{X}$, each element of $\mathbf{R}_{\ell,k}$ and $\mathbf{T}_{\ell,k}$ is defined as
\begin{subequations}
\begin{equation}
\big[\mathbf{R}_{\ell,k}\big]_{q,m}\!\! =\! \beta_{\ell,k}\!\!\int\! e^{2\pi j d (q-m) \sin (\varphi^{\mathrm{Rx}}_{\ell,k}) } f (\varphi^{\mathrm{Rx}}_{\ell,k}) \;{\rm d} \varphi^{\mathrm{Rx}}_{\ell,k},
\end{equation}
\begin{equation}
\big[\mathbf{T}_{\ell,k}\big]_{q,m}\!\! =\! \beta_{\ell,k}\!\!\int\! e^{2\pi j d (q-m) \sin (\varphi^{\mathrm{Tx}}_{\ell,k}) } f (\varphi^{\mathrm{Tx}}_{\ell,k}) \;{\rm d} \varphi^{\mathrm{Tx}}_{\ell,k},
\end{equation}
where $\beta_{\ell,k}$ captures the total average gain of the multipath components of the path between the $\ell$-th \ac{AP} and the $k$-th \ac{UE}, and $d$ is the antenna spacing given in wavelengths, which is assumed to be the same in all antennas, both at \acp{AP} and \acp{UE}.
\end{subequations}

{
When a \ac{BDRIS} is present, the beamformer is designed over the equivalent \ac{AP}-\ac{UE} channel that combines the direct propagation component and the reflected \ac{BDRIS}-assisted component.
Let $\bHTX{\ell}\in\mathbb{C}^{N_{\mathrm{RIS}}\times N_t}$ denote the channel from the $\ell$-th \ac{AP} to the \ac{BDRIS}, let $\bHRX{k}\in\mathbb{C}^{M\times N_{\mathrm{RIS}}}$ denote the channel from the \ac{BDRIS} to the $k$-th \ac{UE}, and let $\bTheta\in\mathbb{C}^{N_{\mathrm{RIS}}\times N_{\mathrm{RIS}}}$ denote the \ac{BDRIS} response matrix.
The equivalent channel between the $\ell$-th \ac{AP} and the $k$-th \ac{UE} is then defined as
\begin{equation}
  \vspace{-0.5ex}
\label{eq:bdris_equiv_ap_channel}
\bHeqAP{\ell}{k} \triangleq \mathbf{H}_{\ell,k}+\bHRX{k}\bTheta\bHTX{\ell} \in \mathbb{C}^{M\times N_t}.
\end{equation}
The direct-only channel is recovered by removing the reflected term, whereas the \ac{BDRIS}-assisted simulations use \eqref{eq:bdris_equiv_ap_channel} with the corresponding \ac{AP}-\ac{BDRIS} and \ac{BDRIS}-\ac{UE} channel components generated according to the channel model in \eqref{eq:chan_model}.
}

{
For convenience, we define the equivalent \ac{DL} channel matrix from all \acp{AP} to the $k$-th \ac{UE} as the concatenation
}
\begin{equation}
  \vspace{-0.5ex}
\bHeq{k} \triangleq \big[ \bHeqAP{1}{k},\bHeqAP{2}{k}, \dots, \bHeqAP{L}{k} \big] \in \mathbb{C}^{M \times N}.
\end{equation}

Denoting the \ac{TX} beamforming matrices associated with the $k$-th \ac{UE} by $\mathbf{V}_{k} \in \mathbb{C}^{N \times M} \in \mathbb{C}^{M \times M}$, respectively, the complex baseband received signal at the $k$-th \ac{UE} can be described as
\begin{equation}
  \vspace{-0.5ex}
\mathbf{y}_k = \overbrace{\,\bHeq{k} \mathbf{V}_k \mathbf{s}_k}^{\text{Intended signal}}\; + \!\!\!\!\!\overbrace{\!\!\!\!\!\sum_{k' \in \mathcal{K}\backslash \{k\}} \!\!\!\!\!\! \bHeq{k} \mathbf{V}_{k'} \mathbf{s}_{k'}}^{\text{Downlink inter-user interference}} \!\! + \!\! \overbrace{\, \mathbf{n}_k}^{\text{Colored noise}},
\label{eq:sig_model}
\end{equation}
where $\mathbf{s}_k\in \mathbb{E}^{M \times 1}$ is the unitary average power signal intended to the $k$-th \ac{UE}, and $\mathbf{n}_k \in \mathbb{C}^{M \times 1} \sim \mathcal{CN}(0,\sigma^2\bm{I}) $ is the circularly symmetric complex-valued \ac{AWGN} at the $k$-th \ac{UE}.

\vspace{-2ex}
\subsection{Problem Statement}
\label{subsec:problemstatement}

\begin{subequations}

From equation \eqref{eq:sig_model}, it follows that the intended \ac{DL} achievable rate $\eta_k^\mathrm{I}$ of the $k$-th \ac{UE} is given by
\vspace{-0.5ex}
\begin{equation}
\eta_k^\mathrm{I} = \log_2 \big|\mathbf{I}_{M} + \bHeq{k} \mathbf{F}_k \bHeq{k}^\mathrm{H} \mathbf{E}_k^{-1} \big|,
\label{eq:eta_I}
\vspace{-0.5ex}
\end{equation}
where
\begin{equation}
\vspace{-0.5ex}
\mathbf{E}_k \triangleq \!\!\!\!\!\! \sum\limits_{k'\in \mathcal{K}\backslash \{k\}} \!\!\!\!\!\! \bHeq{k} \mathbf{F}_{k'} \bHeq{k}^\mathrm{H} + \sigma^2  \mathbf{I}_M, 
\label{eq:Ek}
\vspace{-0.5ex}
\end{equation}
describes the power of the interference-plus-noise affecting the $k$-th user, with
\begin{equation}
\vspace{-0.5ex}
\label{eq:F}
\mathbf{F}_k \triangleq \mathbf{V}_k \mathbf{V}_k^\mathrm{H},
\vspace{-0.5ex}
\end{equation}
denoting the \ac{PSD} Gramian matrices of the \ac{DL}-\ac{TX} beamformer $\mathbf{V}_{k}$.

\end{subequations}

It also follows from equation \eqref{eq:sig_model} that the rate of information $\eta_{k, e}^\mathrm{L}$ intended for the $k$-th \ac{UE}, which under ideal conditions can be decoded by the $e$-th \ac{UE}, under the assumption that the latter performs perfect \ac{SIC} of its own intended signal is given by
\begin{subequations}
\vspace{-0.5ex}
\begin{equation}
\eta_{k, e}^\mathrm{L} \triangleq \log_2 \big|\mathbf{I}_{M} + \bHeq{e} \mathbf{F}_k \bHeq{e}^\mathrm{H} \mathbf{E}_{k,e}^{-1} \big|,
\label{eq:eta_L}
\vspace{-0.5ex}
\end{equation}
with
\begin{equation}
\vspace{-0.5ex}
\mathbf{E}_{k,e} \triangleq \!\!\!\!\!\!\! \sum\limits_{k'\in \mathcal{K}\backslash \{k, e\}} \!\!\!\!\!\!\! \bHeq{e} \mathbf{F}_{k'} \bHeq{e}^\mathrm{H} + \sigma^2 \mathbf{I}_M,
\label{eq:Eke}
\end{equation}
describing the power of the interference-plus-noise at the $e$-th \ac{UE} when decoding information intended for the $k$-th \ac{UE}.

\end{subequations}

From the achievable communication rate $\eta_k^\mathrm{I}$ and the leaked communication rate $\eta_{k, e}^\mathrm{L}$, respectively given in equations \eqref{eq:eta_I} and \eqref{eq:eta_L}, the minimum leakage-free (secrecy) achievable rate of the $k$-th user, under the assumption that the eavesdroppers are the other users acting in an opportunistic, self-serving and non-colluding manner, is given by
\begin{subequations}
\label{eq:priv_rate}
\begin{equation}
\label{eq:priv_rateOriginal}
\eta_k \triangleq \min_{e \neq k}\; (\eta_k^\mathrm{I} - \eta_{k,e}^\mathrm{L})^+,\; \text{with} \; (e, k) \, \in \, \mathcal{K},
\end{equation}
which, given that the intended rate $\eta_k^\mathrm{I}$ is constant to the $\min$ operator, and that the minimum of the negated leakage rates $\eta_{k,e}^\mathrm{L}$ is equivalent to a negated maximum, can be relaxed into
\begin{equation}
\label{eq:priv_rateRelaxed}
\eta_k = \eta_k^\mathrm{I} - \max_{e \neq k}(\eta_{k,e}^\mathrm{L}),\; \text{with} \; (e, k) \, \in \, \mathcal{K},
\end{equation}
under the understanding that the link of a $k$-th user whose channel realization yields\footnote{\setlength{\baselineskip}{9pt}Although such ill condition does occur occasionally with the \ac{MMSE} \ac{TX}-\ac{BF}, it has not been observed in our simulations when the proposed \ac{SecLM}-\ac{BF} schemes of Section \ref{sec:priv_prop} is employed, which corroborates the accuracy of the relaxation of equation \eqref{eq:priv_rateOriginal} into \eqref{eq:priv_rateRelaxed}.} $\eta_k^\mathrm{I} < \eta_{k,e}^\mathrm{L}$, for some $e$, is fundamentally insecure; thus, the design of \ac{SecLM}-\ac{BF} for such a user is out of the scope of interest of the article.

\end{subequations}
As argued in Section \ref{sec:introduction}, the definition of the minimum  secrecy (or \emph{private}) rate given in equation \eqref{eq:priv_rate}, where the possible eavesdroppers to a $k$-th user are the other $K-1$ users $e\neq k \in \mathcal{K}$ of the system itself, is motivated in the interest in scenarios of practical relevance, where the members of the system will not sacrifice their own \ac{QoS} in order to eavesdrop, thus fulfilling the assumption that the potential eavesdroppers' channels $\bHeq{e}$ are known to the \acp{AP}.

From all the above, we can concisely state that the goal of the article is to design low-complexity \ac{SecLM}-\ac{BF} schemes to maximize $\eta_k$ considering also power limits, which can be expressed mathematically as
\begin{subequations}
\label{eq:ProblemStatement}
\begin{IEEEeqnarray}{L"L}
\underset{\mathbf{V}_k}{\mathrm{maximize}} & \sum_{k\in\mathcal{K}} \eta_k^\mathrm{I} - \max_{ e \neq k}(\eta_{k,e}^\mathrm{L}), \\ 
\mathrm{subject\  to} &  \sum_{k\in\mathcal{K}}\|\mathbf{V}_k\|^2_\mathrm{F} \leq P_{\mathrm{max}},
\end{IEEEeqnarray}
where $P_\mathrm{max}$ denotes the maximum available \ac{TX} power.
\end{subequations}

This motivates us to consider, rather than relaxing $\max(\cdot)$ into a sum and then applying the affine upper-bound, to instead employ the aforementioned ``retrofitting'' approach to relax the $\max(\cdot)$ operator directly, as follows.
To that end, let us first identify, for each $k$-th user, the eavesdropper that achieves the highest leakage rate under the set of matrices $\big\{\mathbf{F}_{k}^{(i^\text{SDP}\!\! -1)}, \forall k \big\}$ obtained at the $(i^\text{SDP}\!\!-1)$-th iteration, denoted
\begin{equation}
\label{eq:e_k}
\tilde{e}_k \triangleq \Big\{e \; \big| \; {\mathrm{arg} \max_{e\neq k}} \; \eta_{k,e}^\mathrm{L}\Big(\big\{\mathbf{F}_k^{(i^\text{SDP}\!\!-1)}, \forall k \big\} \Big) \Big\},
\vspace{-1ex}
\end{equation}
where we indicate in the notation that the leakage rate $\eta_{k,e}^\mathrm{L}$ at the $i$-th iteration is computed via equation \eqref{eq:eta_L}, using the solution $\big\{\mathbf{F}_{k}^{(i^\text{SDP}\!\! -1)}, \forall k \big\}$ obtained in the previous iteration.

Then, the following problem can be formulated as
\begin{subequations}
\label{eq:priv_secbf_sdp_prop}
\begin{IEEEeqnarray}{L"L}
\label{eq:priv_secbf_sdp_prop_obj}
\underset{\mathbf{V}_k}{\mathrm {maximize}} & \sum_{k\in\mathcal{K}} \eta_k^\mathrm{I} - \!\! \sum_{k,\tilde{e}_k\in\mathcal{K}} \!\! \eta_{k,\tilde{e}_k}^\mathrm{L}\\
\mathrm{subject\  to} &  \sum_{k\in\mathcal{K}}\|\mathbf{V}_k\|^2_\mathrm{F} \leq P_{\mathrm{max}}.
\end{IEEEeqnarray}
\end{subequations}

{
In the \ac{BDRIS}-assisted setting, the equivalent channels in \eqref{eq:priv_secbf_sdp_prop} are evaluated for a fixed scattering matrix $\bTheta$ during the \ac{BF} update; the scattering matrix is optimized separately in the alternating \ac{BDRIS} step following \cite{FidanovskiArx2026, SandovalArx2026}.
}

Although the optimal solution to the problem is independent of the \ac{RX}-\ac{BF}, in practice, they affect the \ac{BER} experienced by users. Therefore, for completeness, we adopt the MMSE-based receive beamforming solution as in \cite{FritzscheWCNC2013}
\begin{equation}
\label{eq:RXbf_mmse}
\mathbf{U}_{k}=\mathbf{V}_{k}^{\mathrm{H}} \bHeq{k}^{\mathrm{H}}\left(\bHeq{k} \mathbf{F} \bHeq{k}^{\mathrm{H}}+\sigma^{2} \mathbf{I}_{N}\right)^{-1}, 
\end{equation}
with $\mathbf{F} = \mathbf{V}\mathbf{V}^{\mathrm{H}} \in \mathbb{C}^{N \times N}.$

\vspace{-1ex}
\section{SotA: Leakage Minimization Beamformers}
\label{sec:leak_min_sota}
\Ac{SotA} leakage minimization beamforming methods for \ac{CF-mMIMO} systems focus on securing \ac{DL} transmission when legitimate \acp{UE} may opportunistically eavesdrop on each other. 
The main goal is to design \ac{TX} beamformers that maximize the secrecy rate for each user, i.e., the difference between the intended rate and the maximum leakage rate to any other user, subject to a total power constraint. 
These approaches formulate the secrecy rate maximization as a non-convex optimization problem and employ convex relaxation or approximation techniques to obtain tractable solutions.

To render the problem tractable, it is commonly relaxed using techniques such as \ac{SDP} or approximated using the affine bounds. 
These approaches exploit the mathematical structure of the achievable rate and leakage rate expressions, which are characterized by log-determinant functions of the \ac{SINR} covariance matrices. 
Through iterative updates of the beamforming matrices and auxiliary variables, these algorithms converge to global optimal solutions that balance communication performance and secrecy.

The following summarizes the principal steps and mathematical formulations underlying the \ac{SotA} leakage minimization beamforming methodology.
The $k$-th term of the objective function, as derived in equation \eqref{eq:priv_secbf_sdp_prop_obj} from \cite{Sandoval23}, represents the secrecy rate for user $k$ and is expressed as
\begin{align}
\eta_k^\mathrm{I} - \eta_{k,\tilde{e}_k}^\mathrm{L} &= \log_2 \left| \mathbf{E}_k + \bHeq{k} \mathbf{F}_k \bHeq{k}^\mathrm{H} \right| - \log_2 \left| \mathbf{E}_k \right|\\
&\quad - \log_2 \underbrace{\left| \mathbf{E}_{k,\tilde{e}_k} + \bHeq{\tilde{e}_k} \mathbf{F}_k \bHeq{\tilde{e}_k}^\mathrm{H} \right|}_{= \mathbf{E}_{\tilde{e}_k}} + \log_2 \left| \mathbf{E}_{k,\tilde{e}_k} \right|, \notag
\end{align}
which in turn can be lower bounded by
\vspace{-1ex}
\begin{equation}
\eta_k^\mathrm{I}\!-\!\eta_{k,\tilde{e}_k}^\mathrm{L}\!\geq\!\hat{\eta}_k^\mathrm{I}\!+\!\hat{\eta}_{k,\tilde{e}_k}^\mathrm{L}\!+\!\overbrace{\log_2 \left| \mathbf{D}_k \right|\!+\!\log_2 \left|\mathbf{D}_{\tilde{e}_k} \right|}^{\text{independent of } \mathbf{F}_k} +\!\overbrace{\frac{2M}{\ln(2)}}^{\text{constant}}.\!
\end{equation}

\begin{subequations}
\label{eq:priv_rm_sdp_final_opt}
Finally, dropping the terms independent on $\mathbf{F}_k$, as it was done in \cite{Sandoval23}, the following concave equivalent of problem \eqref{eq:priv_secbf_sdp_prop} is obtained
\vspace{-0.5ex}
\begin{IEEEeqnarray}{L"L}
\label{eq:priv_rm_sdp_final_opt_obj?}
\underset{\mathbf{F}_k \succeq 0}{\mathrm {maximize}} & \sum_{k\in\mathcal{K}} \hat{\eta}_k^\mathrm{I} + \!\! \sum_{k,\tilde{e}_k\in\mathcal{K}} \!\! \hat{\eta}_{k,\tilde{e}_k}^\mathrm{L} \\
\mathrm{subject\  to} &  \sum_{k\in\mathcal{K}} \mathrm{Tr} (\mathbf{F}_k)\leq P_{\mathrm{max}},
\end{IEEEeqnarray}
\end{subequations}
where we have highlighted the terms that are dependent on the optimization variable $\mathbf{F}_k$, and the affine rate lower-bounds $\hat{\eta}_k^\mathrm{I}$ and $\hat{\eta}_{k,\tilde{e}_k}^\mathrm{L}$ are respectively defined as
\vspace{-0.5ex}
\begingroup
\renewcommand{\theHequation}{secrecybound.\arabic{equation}}
\begin{IEEEeqnarray}{c}
\label{eq:secrecybound}
\hat{\eta}_k^\mathrm{I} \triangleq \log_2 \big| \mathbf{E}_k + \bHeq{k} \mathbf{F}_k \bHeq{k}^\mathrm{H} \big| - \dfrac{\mathrm{Tr}(\mathbf{D}_k \mathbf{E}_k)}{\ln(2)}, \\
\hat{\eta}_{k,\tilde{e}_k}^\mathrm{L} \triangleq \log_2 \big| \mathbf{E}_{k,\tilde{e}_k} \big| - \dfrac{\mathrm{Tr}(\mathbf{D}_{\tilde{e}_k} \mathbf{E}_{\tilde{e}_k})}{\ln(2)}.
\end{IEEEeqnarray}
\endgroup

Then, each subsequent $i^\text{SDP}\!\!$-th iteration of the solver is executed using the constant matrices $\mathbf{D}_k,\,\forall\,k$ constructed using the matrices $\mathbf{F}_k$ obtained at the $(i^\text{SDP}\!\!-1)$-th iteration, which shall hereafter be denoted $\mathbf{F}_k^{(i^\text{SDP}\!\!-1)}$, yields \cite{Omid_TWX2022}
\vspace{-1ex}
\begin{equation}
\mathbf{D}_k\! \triangleq\!  \Bigg[\overbrace{\sum\limits_{k'\in \mathcal{K}\backslash \{k\}} \!\!\!\!\!\! \bHeq{k} \mathbf{F}_{k'}^{(i^\text{SDP}\!\!-1)} \bHeq{k}^\mathrm{H} \! +\! \sigma^2 \mathbf{I}_M}^{\triangleq \mathbf{E}_k^{(i^\text{SDP}\!\!-1)}}\Bigg]^{\!-1}\!\!\!\!\!\!,
\label{eq:Gk}
\vspace{-1ex}
\end{equation}
where we have implicitly defined the quantities $\mathbf{E}_k^{(i^\text{SDP} \hspace{-0.5ex} -1)}$, which capture the interference-plus-noise that would affect the $k$-th \ac{UE} if the \ac{BF} vectors obtained in the $(i^\text{SDP}\!\!-1)$-th iteration were employed.

These serve as the basis for the iterative construction of the matrix $\mathbf{D}_k$, which results from applying the first order Taylor approximation of $\log_2 \big| \mathbf{E}_k^{-1} \big|$, at a point $\mathbf{D}_k^{-1}$
\vspace{-2.5ex}
\begin{equation}
-\log_{2}\left|\mathbf{E}_{k}\right| \geq \log_{2}\left|\mathbf{D}_{k}\right| - \frac{\operatorname{Tr}\left(\mathbf{D}_{k} \mathbf{E}_{k}\right)}{\ln (2)}+\frac{M}{\ln (2)}.
\vspace{-1.5ex}
\end{equation}

With this in mind, the problem in equation \eqref{eq:priv_rm_sdp_final_opt} can be solved using interior point methods \cite{diamond2016cvxpy, mosek, MyListOfPapers:SturmWhitePaper2001, MyListOfPapers:BoydBook2004}, and is summarized above in Algorithm \ref{alg:priv_rm_alg_sdp}.

\vspace{-3ex}
\section{Proposed Leakage-Minimizing Beamformer}
\label{sec:priv_prop}
\vspace{-0.5ex}

In this section, we introduce a novel \ac{SecLM}-\ac{BF} scheme leveraging the \ac{FP} framework~\cite{shen2019} to efficiently address the non-convex optimization problem in~\eqref{eq:priv_secbf_sdp_prop}. Unlike the computationally intensive SotA \ac{SDP}-based approach described in Section~\ref{sec:leak_min_sota}, the proposed method achieves significant reductions in complexity while maintaining secrecy performance comparable to recent \ac{FP}-based beamforming designs~\cite{shen2018-I}. This enables scalable and practical beamforming for large-scale systems without compromising solution quality.

\vspace{-3ex}
\subsection{Approximation into Concave Convex Problem}

In this subsection, first the optimization problem \eqref{eq:priv_secbf_sdp_prop} is approximated into a form of concave convex problem with a procedure of the matrix \ac{FP}\cite{shen2019}.

The objective function in optimization problem \eqref{eq:priv_secbf_sdp_prop} is characterized by intended communication rate $\eta_k^\mathrm{I}$ minus leakage communication rate $\eta_{k,e}^\mathrm{L}$, and both of the rate function is a form of $\log_2$ of determinant, which makes the optimization problem non-concave.

\begin{algorithm}[!t]
\caption{\!\!:\, {\bf SotA Method} - \ac{DL}-\ac{SecLM}-\ac{BF} via \ac{SDP}-\ac{TX} and \ac{MMSE}-\ac{RX} \acp{BF}}
\begin{algorithmic}[1]
\renewcommand{\algorithmicrequire}{\textbf{Input:}}
\renewcommand{\algorithmicensure}{\textbf{Output:}}
\Statex{\hspace{-3.5ex}\textbf{Internal Parameters:}} Maximum \ac{TX} power $P_\mathrm{max}$, channel 
\Statex \hspace{-3.5ex}matrices $\bHeq{k}$, noise power $\sigma^2$, convergence tolerance $\epsilon$, and
\Statex \hspace{-3.5ex}maximum number of iterations $i^\text{SDP}_{\max}$
\Statex{\hspace{-3.5ex}\textbf{Output:}} \ac{TX} and \ac{RX} \ac{DL}-\ac{BF} matrices $\mathbf{V}_k \text{ and } \mathbf{U}_k, \forall k$
\vspace{-1.5ex}
\Statex \hspace{-3.5ex}\hrulefill
\Statex{\hspace{-3.5ex}\textbf{Initialization:}} Obtain $\mathbf{V}_k^{(0)}$ as the output of Algorithm 3 in \cite{Sandoval23}, and $\mathbf{F}_k^{(0)}$ via \eqref{eq:F}, $\forall k$
\vspace{-1.5ex}
\Statex \hspace{-3.5ex}\hrulefill
\Repeat 
\State Increment iteration index $i^\text{SDP}$ by 1
\State Compute $\mathbf{D}_k,\forall k$ as in eq \eqref{eq:Gk}
\State Compute $\tilde{e}_k,\forall k$ as in eq \eqref{eq:e_k}
\State Obtain $\mathbf{F}_k^{(i^\text{SDP})}$, $\forall k$ by solving \eqref{eq:priv_rm_sdp_final_opt}
\Until {$\big\| \mathbf{F}_k^{(i^\text{SDP})} - \mathbf{F}_k^{(i^\text{SDP}\!\!-1)} \big\|_\text{F} < \epsilon, \forall k$} \textbf{or} $i^\text{SDP} \!\! = i^\text{SDP}_{\max}$
\State Extract $\mathbf{V}_k$ as the square root of $\mathbf{F}^{(i^\text{SDP})}_k, \forall k$
\State Compute $\mathbf{U}_k, \forall k$ via \eqref{eq:RXbf_mmse}
\State \textbf{return} $\mathbf{U}_k, \forall k$ and $\mathbf{V}_k, \forall k$
\end{algorithmic}
\label{alg:priv_rm_alg_sdp}
\end{algorithm}

First, we approximate the optimization without operation of the $\log$ with the \ac{LDT} \cite[Sec. III-B Th.2]{shen2019} as
\begin{subequations}
\begin{IEEEeqnarray}{rCl}
\label{eq:etaILDT}
\check{\eta}_k^\mathrm{I} & = & \overbrace{\log_2\! \big| \mathbf{I}_{M}\! + \!\mathbf{\Gamma}_k^\mathrm{I} \big|\! -\! \mathrm{Tr}\big( \mathbf{\Gamma}_k^\mathrm{I} \big)}^{\text{constant with respect to $\mathbf{V}_{k}$}} \\
& & + \mathrm{Tr}\Big(\hspace{-4ex}\underbrace{\big(\mathbf{I}_{M}\! +\! \mathbf{\Gamma}_k^\mathrm{I} \big)}_{\text{constant with respect to $\mathbf{V}_{k}$}} \hspace{-4ex} \bHeq{k} \mathbf{V}_{\!k} {\mathbf{M}_k^\mathrm{I}}^{-1}  \mathbf{V}_{k}^\mathrm{H} \bHeq{k}^\mathrm{H} \Big),\nonumber
\end{IEEEeqnarray}
\begin{IEEEeqnarray}{rCl}
\label{eq:etaLLDT}
\check{\eta}_{k,\tilde{e}_k}^\mathrm{L} & = & \overbrace{\log_2\! \big| \mathbf{I}_{M}\! + \!\mathbf{\Gamma}_{k,\tilde{e}_k}^\mathrm{L} \big|\! -\! \mathrm{Tr}\big( \mathbf{\Gamma}_{k,\tilde{e}_k}^\mathrm{L} \big)}^{\text{constant with respect to $\mathbf{V}_{k}$}} \\
& & + \mathrm{Tr}\Big(\hspace{-3.5ex} \underbrace{\big(\mathbf{I}_{M}\! +\! \mathbf{\Gamma}_{k,\tilde{e}_k}^\mathrm{L} \big)}_{\text{constant with respect to $\mathbf{V}_{k}$}} \hspace{-3.5ex} \bHeq{\tilde{e}_k} \mathbf{V}_{\!k} {\mathbf{M}_{k,\tilde{e}_k}^\mathrm{L}}^{-1}  \mathbf{V}_{k}^\mathrm{H} \bHeq{\tilde{e}_k}^\mathrm{H} \Big),\nonumber
\end{IEEEeqnarray}
\end{subequations}
where $\mathbf{\Gamma}_k^\mathrm{I}$ $\mathbf{\Gamma}_{k,\tilde{e}_k}^\mathrm{L}$ are the matrix Lagrange multipliers, given by
\begin{subequations}
\label{eq:fLDT}
\begin{equation}
\label{eq:GammaI}
\mathbf{\Gamma}_k^\mathrm{I} \triangleq \bHeq{k} \big[\mathbf{V}_{k} \mathbf{E}_k^{-1} \mathbf{V}^\mathrm{H}_{k}\big]^\mathrm{(i^\text{FP}\!\!-1)} \bHeq{k}^\mathrm{H},
\end{equation}
\begin{equation}
\label{eq:GammaL}
\mathbf{\Gamma}_{k,\tilde{e}_k}^\mathrm{L} \triangleq \bHeq{\tilde{e}_k} \big[\mathbf{V}_{k} \mathbf{E}_{k,\tilde{e}_k}^{-1} \mathbf{V}^\mathrm{H}_{k}\big]^\mathrm{(i^\text{FP}\!\!-1)} \bHeq{\tilde{e}_k}^\mathrm{H},
\end{equation}
and we have, for convenience of notation, introduced the auxiliary matrices $\mathbf{M}_{k}^\mathrm{I}$, $\mathbf{M}_{k,e}^\mathrm{L}$ defined as
\begin{eqnarray}
\label{eq:Mk}
&\mathbf{M}_{k}^\mathrm{I} \triangleq \sum\limits_{k'\in \mathcal{K}}\!\! \bHeq{k} \mathbf{V}_{k'} \mathbf{V}_{k'}^\mathrm{H} \bHeq{k}^\mathrm{H} + \sigma^2 \mathbf{I}_M,&\\
\label{eq:MkL}
&\mathbf{M}_{k,\tilde{e}_k}^\mathrm{L} \triangleq \sum\limits_{k'\in \mathcal{K}\backslash\{\tilde{e}_k\}}\!\! \bHeq{\tilde{e}_k} \mathbf{V}_{k'} \mathbf{V}_{k'}^\mathrm{H} \bHeq{\tilde{e}_k}^\mathrm{H} + \sigma^2 \mathbf{I}_M.&
\end{eqnarray}
\end{subequations}

As emphasized in equation \eqref{eq:fLDT}, the Lagrange multiplier matrices $\mathbf{\Gamma}_k^\mathrm{I}$, $\mathbf{\Gamma}_{k,\tilde{e}_k}^\mathrm{L}$ are constant with respect to the optimization variables $\mathbf{V}_{k}$.
In anticipation of the algorithmic procedure to be described in the sequel, we have therefore utilized the same notation introduced in the previous subsection, indicating in equation \eqref{eq:GammaI} and \eqref{eq:GammaL} that the term $\big[\mathbf{V}_{k} \mathbf{E}_k^{-1} \mathbf{V}^\mathrm{H}_{k}\big]^\mathrm{(i^\text{FP} \!\!-1)}$ and $\big[\mathbf{V}_{k} \mathbf{E}_{k,\tilde{e}_k}^{-1} \mathbf{V}^\mathrm{H}_{k}\big]^\mathrm{(i^\text{FP} \!\!-1)}$ are to be computed for a given ``point,'' that is, using the solution obtained at the previous iteration.

Notice, however, that the expression in equation \eqref{eq:fLDT} is not concave, due to the ratio on the variable $\mathbf{V}_{k}$ embedded in the last term of $\check{\eta}_k^\mathrm{I}$ and $\check{\eta}_{k,\tilde{e}_k}^\mathrm{L}$.

This can be circumvented by the \ac{QT} described in \cite[Sec. III-B, Th.1]{shen2019}, which applied onto equation \eqref{eq:fLDT} yields
\begin{subequations}
\begin{IEEEeqnarray}{rCl}
\label{eq:fQTI}
\breve{\eta}_k^\mathrm{I} & = & \log_2 \big| \mathbf{I}_{M} \! + \! \mathbf{\Gamma}_k^\mathrm{I} \big| \! - \! \mathrm{Tr}\big( \mathbf{\Gamma}_k^\mathrm{I} \big) \\[-1ex]
& & + \mathrm{Tr}\Big( \! \big( \mathbf{I}_{M}\! +\! \mathbf{\Gamma}_k^\mathrm{I} \big)  \big( 2 \Re \{  \mathbf{V}_{k}^\mathrm{H} \bHeq{k}^\mathrm{H} \mathbf{Y}_k^\mathrm{I} \}\! - \! {\mathbf{Y}_k^\mathrm{I}}^\mathrm{H} \mathbf{M}_k^\mathrm{I} \mathbf{Y}_k^\mathrm{I}\big) \Big),\nonumber
\end{IEEEeqnarray}
\begin{IEEEeqnarray}{rCl}
\label{eq:fQTL}
\hspace{-8ex}\breve{\eta}_{k,\tilde{e}_k}^\mathrm{L} & = & \log_2 \big| \mathbf{I}_{M} \! + \! \mathbf{\Gamma}_{k,\tilde{e}_k}^\mathrm{L} \big| \! - \! \mathrm{Tr}\big( \mathbf{\Gamma}_{k,\tilde{e}_k}^\mathrm{L} \big) \\[-1ex]
& & + \mathrm{Tr}\Big( \! \big( \mathbf{I}_{M}\! +\! \mathbf{\Gamma}_{k,\tilde{e}_k}^\mathrm{L} \big)  \big( 2 \Re \{  \mathbf{V}_{k}^\mathrm{H} \bHeq{\tilde{e}_k}^\mathrm{H} {\mathbf{Y}_{k,\tilde{e}_k}^\mathrm{L}} \}  \nonumber \\[-1ex]
& & - {\mathbf{Y}_{k,\tilde{e}_k}^\mathrm{L}}^\mathrm{\!\!\!\!H} \mathbf{M}_{k,\tilde{e}_k}^\mathrm{L} {\mathbf{Y}_{k,\tilde{e}_k}^\mathrm{L}} \big) \Big),\nonumber
\end{IEEEeqnarray}
\noindent where
\begin{equation}
\mathbf{Y}_k^\mathrm{I} = \raisebox{-3pt}{\Bigg[}\!\!\sum\limits_{k'\in \mathcal{K}}\!\!  \bHeq{k} \mathbf{F}_{k'}^{\!\raisebox{4pt}{\scriptsize $(i^\text{FP}\!\!-\!1)$}} \bHeq{k}^\mathrm{H} \! +\! \sigma^2 \mathbf{I}_M \!\raisebox{-3.5pt}{\Bigg]}^{\!-1}\!\!\!\!\!\! \bHeq{k}\! \mathbf{V}_{k}^{(i^\text{FP}\!\!-\!1)}\!\!\!\!\!\!\!\!,
\label{eq:YI}
\end{equation}
\vspace{-1ex}
\begin{equation}
\mathbf{Y}_{k,\tilde{e}_k}^\mathrm{L} \! = \! \raisebox{-3pt}{\Bigg[}\!\!\sum\limits_{k'\in \mathcal{K} \backslash\{\tilde{e}_k\} }\!\!\!\!\!\!\!  \bHeq{\tilde{e}_k} \mathbf{F}_{k'}^{\!\raisebox{4pt}{\scriptsize $(i^\text{FP}\!\!-\!1)$}} \bHeq{\tilde{e}_k}^{\!\!\!\mathrm{H}} \! +\! \sigma^2 \mathbf{I}_M \!\raisebox{-3.5pt}{\Bigg]}^{\!-1}\!\!\!\!\!\! \bHeq{\tilde{e}_k}\! \!\mathbf{V}_{k}^{(i^\text{FP}\!\!-\!1)}\!\!\!\!\!\!\!\!.
\label{eq:YL}
\end{equation}
\end{subequations}

\begin{subequations}
\label{eq:secbf_FP}

Finally, the optimization problem for secrecy rate maximization can be relaxed into a concave convex problem as
\begin{IEEEeqnarray}{L"L}
\underset{\mathbf{V}_k}{\mathrm {maximize}} & \underbrace{ \sum_{k\in\mathcal{K}} \breve{\eta}_k^\mathrm{I}}_{\text{Concave}} \hspace{3ex} \underbrace{  -  \sum_{k,\tilde{e}_k\in\mathcal{K}}  \breve{\eta}_{k,\tilde{e}_k}^\mathrm{L}  }_{\text{Convex}} \label{eq:secbf_FP_obj}\\
\mathrm{subject\  to} &  \sum_{k\in\mathcal{K}}\|\mathbf{V}_k\|^2_\mathrm{F} \leq P_{\mathrm{max}},
\end{IEEEeqnarray}
\end{subequations}

The resulting problem in \eqref{eq:secbf_FP} is a difference of concave and convex program, which remains non-convex and requires further processing to enable efficient optimization.

\vspace{-1ex}
\subsection{Relaxation via CCP}
To address this challenge, we adopt the CCP~\cite{yuille03}, an iterative method that successively approximates the non-convex problem by linearizing the convex part of the objective at each iteration.
This transforms the original problem into a sequence of tractable convex subproblems that can be efficiently solved.

In general, a differentiable difference-of-convex (DC) program can be relaxed using CCP as follows:
\vspace{-1ex}
\begin{subequations}
\begin{eqnarray}
\underset{\mathbf{X}}{\mathrm{minimize}} \!\!\!&& g_0(\mathbf{X}|\mathbf{X}^{i^\text{CCP}}) \\
\mathrm{subject\ to} \!\!\!&& g_k(\mathbf{X}|\mathbf{X}^{i^\text{CCP}}) \leq 0,\; k \in 
\{1,\dots,K\},
\end{eqnarray}
\end{subequations}
where
\vspace{-1ex}
\begin{eqnarray}
g_k(\mathbf{X}|\mathbf{X}^{i^\text{CCP}}) = f_k(\mathbf{X}) - h_k(\mathbf{X}^{i^\text{CCP}})&& \\
&&\hspace{-32ex} + \mathrm{Tr} \big\{ \nabla h_k(\mathbf{X}^{i^\text{CCP}}) \cdot \mathbf{X} \big\}
- \mathrm{Tr} \big\{ \nabla h_k(\mathbf{X}^{i^\text{CCP}}) \cdot \mathbf{X}^{i^\text{CCP}} \big\}.\nonumber
\end{eqnarray}

Applying CCP to our problem, the convex part of the objective (the sum of leakage terms) is linearized at each iteration, resulting in a surrogate problem
\vspace{-1ex}
\begin{subequations}
\begin{eqnarray}
\label{eq:secFP_CCP}
\underset{\mathbf{V}_k}{\mathrm{maximize}} && \sum_{k\in\mathcal{K}} \breve{{\eta}}_k^\mathrm{I} - \mathcal{L}(\mathbf{V}_k) \\
\label{eq:secFP_CCP_constraint}
\mathrm{subject\  to} && \sum_{k\in\mathcal{K}}\left\| \mathbf{V}_k \right\|_\mathrm{F}^2 \leq P_{\max},
\end{eqnarray}
\end{subequations}
at iteration $i^\text{CCP}$, where gradients are evaluated using the previous quantities $\mathbf{V}_k^{i^\text{CCP}}$, and the linearized leakage becomes
\vspace{-1ex}
\begin{equation}
\mathcal{L}(\{\mathbf{V}_k\}) = \!\!\!\!\! \sum_{k,\tilde{e}_k \in\mathcal{K}} \!\!\!\! \mathrm{Tr} \raisebox{-3pt}{\bigg\{} \nabla_{\!\mathbf{V}_k} \! \raisebox{-3pt}{\bigg[} \! \sum_{k',\tilde{e}_{k'} \in\mathcal{K}} \!\!\!\!\! \breve{{\eta}}_{k',\tilde{e}_{k'}}^\mathrm{L}  \raisebox{-3pt}{\bigg ]} \! \cdot \! \big(\mathbf{V}_k\! -\! \mathbf{V}^{i^\text{CCP}}_k\big) \!  \raisebox{-3pt}{\bigg\}},
\end{equation}
with the complex matrix gradient $\nabla f(\cdot)$ is given by\cite{petersen2012}
\begin{equation}
\nabla f(\mathbf{X}) = \frac{\partial f(\mathbf{X})}{ \partial \Re \mathbf{X}} + j \frac{\partial f(\mathbf{X})}{ \partial \Im \mathbf{X}}.
\end{equation}

Then, the gradient of $\breve{\eta}_{k,\tilde{e}_k}^\mathrm{L}$ from \eqref{eq:secFP_CCP} can be split into
\begin{subequations}
\begin{equation}
\nabla \sum\limits_{k\in \mathcal{K}} \breve{\eta}_{k,\tilde{e}_k}^\mathrm{L} = \nabla \sum\limits_{k\in \mathcal{K}} f_{k,\tilde{e}_k}^\mathrm{lin} - \nabla \sum\limits_{k\in \mathcal{K}} f_{k,\tilde{e}_k}^\mathrm{quad},
\end{equation}
where
\begin{equation}
f_{k,\tilde{e}_k}^\mathrm{lin} \triangleq 2 \mathrm{Tr}\Big\{ \big( \mathbf{I}_{M}\! +\! \mathbf{\Gamma}_{k,\tilde{e}_k}^\mathrm{L} \big) \Re \{  \mathbf{V}_{k}^\mathrm{H} \mathbf{A}_{k,\tilde{e}_k} \} \Big\},
\label{eq:f_lin}
\end{equation}
\begin{equation}
f_{k,\tilde{e}_k}^\mathrm{quad} \triangleq \mathrm{Tr}\Big\{\mathbf{B}_{k,\tilde{e}_k} \!\!\!\! \sum\limits_{k'\in \mathcal{K} \backslash \{k\} } \!\!\!\!\! \mathbf{V}_{k'} \mathbf{V}_{k'}^\mathrm{H} \Big\} ,
\end{equation}
with
\begin{equation}
\label{eq:A_k}
\mathbf{A}_{k,\tilde{e}_k} \triangleq \bHeq{\tilde{e}_k}^\mathrm{H} \mathbf{Y}_{k,\tilde{e}_k}^\mathrm{L},
\end{equation}
\begin{equation}
\label{eq:B_k}
\mathbf{B}_{k,\tilde{e}_k} \triangleq \mathbf{A}_{k,\tilde{e}_k} \big( \mathbf{I}_{M}\! +\! \mathbf{\Gamma}_{k,\tilde{e}_k}^\mathrm{L} \big) \mathbf{A}_{k,\tilde{e}_k}^\mathrm{H}.
\end{equation}
\end{subequations}

Looking only at the part of the gradient affected by some particular $\mathbf{V}_k$ we get
\begin{equation}
\nabla \sum\limits_{k\in \mathcal{K}} f_{k,\tilde{e}_k}^\mathrm{quad} = \sum\limits_{k'\in \mathcal{K}} \mathbf{B}_{k',\tilde{e}_{k'}} \mathbf{V}_k, \; \quad \forall k.
\end{equation}

To address the linear part, it is first transformed under the assumption that the matrices inside the real operation in \eqref{eq:f_lin} are nearly diagonal:
\begin{equation}
f_{k,\tilde{e}_k}^\mathrm{lin} = \mathrm{Tr}\Big\{ \big( \mathbf{I}_{M}\! + \! \mathbf{\Gamma}_{k,\tilde{e}_k}^\mathrm{L} \big) \big( \mathbf{V}_{k}^\mathrm{H} \mathbf{A}_{k,\tilde{e}_k} + \mathbf{A}_{k,\tilde{e}_k}^\mathrm{H} \mathbf{V}_{k} \big) \Big\},
\end{equation}
and focusing only on the terms in the gradient that depend on a particular $\mathbf{V}_k$, we obtain
\begin{equation}
\nabla \sum\limits_{k\in \mathcal{K}} f_{k,\tilde{e}_k}^\mathrm{lin} = 2 \Re \big\{ \mathbf{A}_{k,\tilde{e}_k} \big( \mathbf{I}_{M}\! + \! \mathbf{\Gamma}_{k,\tilde{e}_k}^\mathrm{L} \big) \big\}, \; \forall k.
\end{equation}

Finally, combining the above results, the gradient of the sum of leakage terms with respect to $\mathbf{V}_k$ is given by
\begin{eqnarray}
\label{eq:gradient_sum_leakage}
\nabla\sum\limits_{k\in \mathcal{K}} \breve{\eta}_{k,\tilde{e}_k}^\mathrm{L}
= 2 \Re \big\{ \bHeq{\tilde{e}_k}^\mathrm{H} \mathbf{Y}_{k,\tilde{e}_k}^\mathrm{L} \big( \mathbf{I}_{M} +  \mathbf{\Gamma}_{k,\tilde{e}_k}^\mathrm{L} \big) \big\}&& \\[-2ex]
&& \hspace{-38ex}
-\sum\limits_{k'\in \mathcal{K}} \bHeq{\tilde{e}_{k'}}^\mathrm{H} \mathbf{Y}_{k',\tilde{e}_{k'}}^\mathrm{L} \big( \mathbf{I}_{M} +\mathbf{\Gamma}_{k',\tilde{e}_{k'}}^\mathrm{L} \big) \mathbf{Y}_{k',\tilde{e}_{k'}}^{\mathrm{L}^{\scriptstyle\mathrm{H}}} \bHeq{\tilde{e}_{k'}} \mathbf{V}_k.\nonumber
\end{eqnarray}

This expression can now be used within the CCP-based iterative optimization to update the beamforming matrices $\mathbf{V}_k$ at each step.
Moreover, equating the partial derivatives to 0 and solving for the beamforming matrices $\mathbf{V}_k$, we can obtain the optimal beamforming matrices $\mathbf{V}_k$ at each iteration of the CCP algorithm as
\begin{eqnarray}
\label{eq:bf_update_ccp}
\mathbf{V}_k = \Big( \mu_k \mathbf{I}_N + \!\! \sum_{k'\in\mathcal{K}}\! \bHeq{k'}^\mathrm{H} \mathbf{Y}_{\!k'} \big( \mathbf{I}_M + \bm{\Gamma}_{\!k'} \big) \mathbf{Y}_{\!k'}^\mathrm{H} \bHeq{k'} \Big)^{-1} && \\
&& \hspace{-35ex} \Big( \bHeq{k}^\mathrm{H} \mathbf{Y}_{\!k}\! \big( \mathbf{I}_M + \bm{\Gamma}_k \big) - \frac{1}{2}\nabla\sum\limits_{k\in \mathcal{K}} \breve{\eta}_{k,\tilde{e}_k}^\mathrm{L}\Big).\nonumber
\end{eqnarray}

\subsection{\texorpdfstring{{CDM-Based Solution of the Beamformer Update}}{CDM-Based Solution of the Beamformer Update}}

{
The inverse in \eqref{eq:bf_update_ccp} is only needed to solve a linear system.
We therefore rewrite the beamformer update in the aforementioned linear form and solve it by row-block \ac{CDM}, which keeps the \ac{FP}/\ac{CCP} formulation unchanged while avoiding an explicit matrix inverse.
The residual is carried with the beamformer during the row updates, which makes the stopping rule and warm-start mechanism explicit.

To obtain this structure, we define the following matrices, namely $\mathbf{G}$, $\mathbf{Q}_k(\mu_k)$, and $\mathbf{C}_k$ as
\begin{subequations}
\label{eq:cdm_defs}
\begin{gather}
\mathbf{G} \triangleq \sum_{k'\in\mathcal{K}} \bHeq{k'}^\mathrm{H} \mathbf{Y}_{k'} \big( \mathbf{I}_M + \bm{\Gamma}_{k'} \big) \mathbf{Y}_{k'}^\mathrm{H} \bHeq{k'}, \label{eq:cdm_g}\\
\mathbf{Q}_k(\mu_k) \triangleq \mu_k \mathbf{I}_N + \mathbf{G}, \label{eq:cdm_q}\\
\mathbf{C}_k \triangleq \bHeq{k}^\mathrm{H} \mathbf{Y}_{k} \big( \mathbf{I}_M + \bm{\Gamma}_{k} \big) - \frac{1}{2}\nabla\sum\limits_{k\in \mathcal{K}} \breve{\eta}_{k,\tilde{e}_k}^\mathrm{L}. \label{eq:cdm_c}
\end{gather}
\end{subequations}

Then \eqref{eq:bf_update_ccp} is equivalent to the linear system
\begin{equation}
\label{eq:cdm_linear_system}
\mathbf{Q}_k(\mu_k)\mathbf{V}_k = \mathbf{C}_k,
\end{equation}

For $\mu_k>0$, the matrix $\mathbf{Q}_k(\mu_k)$ is Hermitian positive definite, and the Gauss--Seidel coordinate-descent iterations converge to the solution of \eqref{eq:cdm_linear_system} when the inner loop is run to convergence.

Equivalently, the \ac{CDM} iterations minimize the quadratic objective
\begin{equation}
\label{eq:cdm_objective}
\min_{\mathbf{V}_k}\; \mathrm{Tr}\!\left\{\mathbf{V}_k^\mathrm{H}\mathbf{Q}_k(\mu_k)\mathbf{V}_k\right\}
- 2 \Re\!\left\{\mathrm{Tr}\!\left(\mathbf{C}_k^\mathrm{H}\mathbf{V}_k\right)\right\},
\end{equation}
whose first-order optimality condition is \eqref{eq:cdm_linear_system}.

For an efficient implementation, we define the residual matrix
\begin{equation}
\label{eq:cdm_residual}
\mathbf{R}_k \triangleq \mathbf{Q}_k(\mu_k)\mathbf{V}_k - \mathbf{C}_k \in \mathbb{C}^{N\times M}.
\end{equation}

During one \ac{CDM} sweep, if all rows of $\mathbf{V}_k$ are fixed except row $j$, the exact row-block minimizer is obtained from
\begin{subequations}
\label{eq:cdm_row_update}
\begin{gather}
\mathbf{V}_k[j,:] \leftarrow \mathbf{V}_k[j,:] + \bm{\delta}_j. \label{eq:cdm_vupdate}
\end{gather}
where the row correction term $\bm{\delta}_j$ is given as
\begin{gather}
\bm{\delta}_j \triangleq -\frac{[\mathbf{R}_k]_{j,:}}{[\mathbf{Q}_k(\mu_k)]_{jj}}
= -\frac{[\mathbf{R}_k]_{j,:}}{[\mathbf{G}]_{jj}+\mu_k}, \label{eq:cdm_delta}
\end{gather}
\end{subequations}

Thus, each row update only requires the current residual row and a scalar denominator.
After applying \eqref{eq:cdm_vupdate}, the residual is corrected without recomputing the full product $\mathbf{Q}_k(\mu_k)\mathbf{V}_k$ as
\begin{equation}
\label{eq:cdm_residual_update}
\mathbf{R}_k \leftarrow \mathbf{R}_k + [\mathbf{Q}_k(\mu_k)]_{:,j}\bm{\delta}_j
= \mathbf{R}_k + \big([\mathbf{G}]_{:,j} + \mu_k\mathbf{e}_j\big)\bm{\delta}_j,
\end{equation}
where $\mathbf{e}_j$ is the $j$-th canonical basis vector.
After each row correction, the residual remains consistent with the current beamformer and is immediately available for the next row update.

Warm starts are used at both the \ac{CCP} and bisection levels.
At a new \ac{CCP} iterate, the previous beamformer is reused as the initial point, namely
\begin{equation}
\label{eq:cdm_warm_ccp}
\mathbf{V}_k^{(0)} \leftarrow \mathbf{V}_k^{(i_{\mathrm{CCP}}-1)}, \qquad
\mathbf{R}_k^{(0)} \leftarrow \mathbf{Q}_k(\mu_k)\mathbf{V}_k^{(0)} - \mathbf{C}_k.
\end{equation}

Inside the bisection loop, if the multiplier changes from $\mu_k^{\mathrm{old}}$ to $\mu_k^{\mathrm{new}} = \mu_k^{\mathrm{old}} + \Delta \mu_k$, only the diagonal shift of $\mathbf{Q}_k(\mu_k)$ changes while $\mathbf{G}$ and $\mathbf{C}_k$ remain fixed.
The residual can therefore be updated as
\begin{equation}
\label{eq:cdm_warm_bisection}
\mathbf{R}_k^{\mathrm{new}} = \mathbf{R}_k^{\mathrm{old}} + \Delta \mu_k \mathbf{V}_k.
\end{equation}

This update lets successive bisection trials continue from the previously computed beamformer and residual.
After the row sweeps stop, the solver returns the last row-wise iterate,
\begin{equation}
  \mathbf{V}_k^{\mathrm{CDM}} \triangleq
  \begin{bmatrix}
  \mathbf{V}_k[1,:]^{\mathrm{final}} \\
  \mathbf{V}_k[2,:]^{\mathrm{final}} \\
  \vdots \\
  \mathbf{V}_k[N,:]^{\mathrm{final}}
  \end{bmatrix}, \label{eq:cdm_final_vk_matrix}
\end{equation}
such that each row is given by
\begin{equation}
  \vspace{-1ex}
\mathbf{V}_k[j,:]^{\mathrm{final}} = \mathbf{V}_k[j,:]^{(0)} + \sum_{t=1}^{T_k^{\mathrm{stop}}}\bm{\delta}_j^{(t)},
\quad j=1,\ldots,N, \label{eq:cdm_final_vk_row}
\end{equation}
where $\bm{\delta}_j^{(t)}$ is the row correction generated by \eqref{eq:cdm_delta} at sweep $t$, and $T_k^{\mathrm{stop}}$ denotes the sweep index at which the stopping rule is met.
If the residual is driven to zero, the returned matrix $\mathbf{V}_k^{\mathrm{CDM}}$ satisfies \eqref{eq:cdm_linear_system} and coincides with the beamformer characterized by \eqref{eq:bf_update_ccp}.
}

\subsection{Algorithm for Proposed BF Design}

The proposed \ac{SecLM}-\ac{BF} algorithm is summarized in Algorithm~\ref{alg:SecLM_FP_CCP}. 
The procedure begins by initializing the beamforming matrices $\mathbf{V}_k^{(0)}$, which are obtained via a \ac{MMSE} solution:
\begin{equation}
\!\mathbf{V}_k^{(0)} \!= \!\sqrt{\frac{P_{\max}}{K}} \frac{\big(\sum_{k'\in\mathcal{K}} \bHeq{k'}^\mathrm{H} \bHeq{k'}\!+\!\sigma^2\mathbf{I}_{N}\big)^{-1} \bHeq{k}^\mathrm{H}}{\Big\|\big(\sum_{k'\in\mathcal{K}} \bHeq{k'}^\mathrm{H} \bHeq{k'}\!+\!\sigma^2\mathbf{I}_{N}\big)^{-1} \bHeq{k}^\mathrm{H}\Big\|_\mathrm{F}}.
\label{eq:bf_mmse}
\end{equation}

\begin{algorithm}[!t]
\caption{\!\!:\, {\bf Proposed} - \ac{SecLM}-\ac{BF} via FP and CCP}
\begin{algorithmic}[1]
\renewcommand{\algorithmicrequire}{\textbf{Input:}}
\renewcommand{\algorithmicensure}{\textbf{Output:}}
\Statex{\hspace{-3.5ex}\textbf{Internal Parameters:}} Channel matrices $\{\bHeq{k}\}_{k=1}^K$, noise power $\sigma^2$, maximum number of FP iterations $i_{\text{FP}}$, maximum number of CCP iterations $i_{\text{CCP}}$, convergence tolerances $\epsilon_{\text{FP}}, \epsilon_{\text{CCP}}$
\Statex{\hspace{-3.5ex}\textbf{Output:}} Beamforming matrices $\{\mathbf{V}_k^\star\}_{k=1}^K$
\vspace{-1.5ex}
\Statex \hspace{-3.5ex}\hrulefill
\Statex{\hspace{-3.5ex}\textbf{Initialization:}} Obtain $\mathbf{V}_k^{(0)}$ from \eqref{eq:bf_mmse}
\vspace{-1.5ex}
\Statex \hspace{-3.5ex}\hrulefill
\For{$n = 1$ to $i_{\text{FP}}$}
\State Compute $\tilde{e}_k,\forall k$ as in eq \eqref{eq:e_k}
\State Compute $\mathbf{M}^\mathrm{I}_{k}$, $\mathbf{\Gamma}_k^\mathrm{I}$, $\mathbf{Y}_k^\mathrm{I}$ via \eqref{eq:Mk}, \eqref{eq:GammaI}, \eqref{eq:YI}
\For{$i = 1$ to $i_{\text{CCP}}$}
  \State Obtain $\mathbf{M}_{k,e}^\mathrm{L}$, $\mathbf{\Gamma}_{k,e}^\mathrm{L}$, $\mathbf{Y}_{k,e}^\mathrm{L}$ via \eqref{eq:MkL}, \eqref{eq:GammaL}, \eqref{eq:YL}
  \State Obtain $\mathbf{A}_{k,\tilde{e}_k}$ and $\mathbf{B}_{k,\tilde{e}_k}$ from \eqref{eq:A_k} and \eqref{eq:B_k}
  \State Compute gradient $\nabla_{\mathbf{V}_k}\!\sum_{k,\tilde e_k}\breve\eta_{k,\tilde e_k}^{\mathrm{L}}$ via \eqref{eq:gradient_sum_leakage}
  \State Obtain $\mathbf{V}_k$ from \eqref{eq:bf_update_ccp} (with bisection on $\mu_k$)
  \If{$\|\mathbf{V}_k^{(i_{\text{CCP}})}-\mathbf{V}_k^{(i_{\text{CCP}}-1)}\|_{\mathrm{F}}  \le\epsilon_{\text{CCP}}$}
    \State \textbf{break}
  \EndIf
\EndFor
\If{$\|\mathbf{V}_k^{(i_{\text{FP}})}-\mathbf{V}_k^{(i_{\text{FP}}-1)}\|_{\mathrm{F}}\le\epsilon_{\text{FP}}$}
  \State \textbf{break}
\EndIf
\EndFor
\State \textbf{return} $\{\mathbf{V}_k^\star\}_{k=1}^K$
\end{algorithmic}
\label{alg:SecLM_FP_CCP}
\end{algorithm}

{
After \eqref{eq:bf_mmse}, Algorithm~\ref{alg:SecLM_FP_CCP} alternates between \ac{FP} auxiliary-variable updates and \ac{CCP} beamformer refinements.
At each \ac{FP} iteration, the useful-rate and leakage-rate terms are recomputed from the current beamformers, after which the \ac{CCP} loop forms a local convex surrogate.
Each surrogate update uses bisection on $\mu_k$ to enforce the transmit-power constraint before the next surrogate point is formed.
The two tolerances act at different levels: $\epsilon_{\text{CCP}}$ monitors consecutive surrogate solutions, whereas $\epsilon_{\text{FP}}$ monitors the full beamformer update after the auxiliary variables are refreshed.
Both criteria are matrix-change tests, so convergence is declared only after the beamformers stabilize rather than after a small objective variation alone.
}

{
The row-wise \ac{CDM} update is summarized in Algorithm~\ref{alg:cdm_bf_update}.
The routine initializes the residual, performs Gauss--Seidel row sweeps using \eqref{eq:cdm_row_update} and \eqref{eq:cdm_residual_update}, stops when the residual tolerance or $T_{\max}$ is reached, and returns the assembled beamformer.
For a fixed value of $\mu_k$, the inverse-based update in \eqref{eq:bf_update_ccp} gives the beamformer solving the current surrogate problem, while the bisection search adjusts $\mu_k$ until the transmit-power constraint is satisfied.
Within Algorithm~\ref{alg:SecLM_FP_CCP}, Algorithm~\ref{alg:cdm_bf_update} replaces only this linear-solve step; the \ac{FP} variables, \ac{CCP} surrogate construction, leakage-gradient evaluation, and bisection logic remain unchanged.
\begin{algorithm}[!t]
\caption{\!\!:\, {\bf Proposed -} CDM-based update}
\begin{algorithmic}[1]
\renewcommand{\algorithmicrequire}{\textbf{Input:}}
\renewcommand{\algorithmicensure}{\textbf{Output:}}
\Statex{\hspace{-3.5ex}\textbf{Internal Parameters:}} $\mathbf{G}$ from \eqref{eq:cdm_g}, $\mathbf{C}_k$ from \eqref{eq:cdm_c}, initial beamformer $\mathbf{V}_k^{(0)}$, multiplier $\mu_k$, maximum number of sweeps $T_{\max}$, tolerance $\epsilon_{\mathrm{CDM}}$
\Statex{\hspace{-3.5ex}\textbf{Output:}} CDM-updated beamformer $\mathbf{V}_k^{\mathrm{CDM}}$
\vspace{-1.5ex}
\Statex \hspace{-3.5ex}\hrulefill
\State Form $\mathbf{Q}_k(\mu_k) = \mu_k\mathbf{I}_N + \mathbf{G}$ via \eqref{eq:cdm_q}
\State Initialize $\mathbf{V}_k \leftarrow \mathbf{V}_k^{(0)}$
\State Initialize $\mathbf{R}_k \leftarrow \mathbf{Q}_k(\mu_k)\mathbf{V}_k - \mathbf{C}_k$ via \eqref{eq:cdm_residual}
\For{$t = 1$ to $T_{\max}$}
  \For{$j = 1$ to $N$}
    \State Compute $\bm{\delta}_j$ via \eqref{eq:cdm_delta}
    \State Update $\mathbf{V}_k[j,:]$ via \eqref{eq:cdm_vupdate}
    \State Update $\mathbf{R}_k$ via \eqref{eq:cdm_residual_update}
  \EndFor
  \If{$\|\mathbf{R}_k\|_\mathrm{F} \le \epsilon_{\mathrm{CDM}}$}
    \State \textbf{break}
  \EndIf
\EndFor
\State Assemble the final row-wise solution $\mathbf{V}_k^{\mathrm{CDM}}$ according to \eqref{eq:cdm_final_vk_matrix} and \eqref{eq:cdm_final_vk_row}
\State \textbf{return} $\mathbf{V}_k^{\mathrm{CDM}}$
\end{algorithmic}
\label{alg:cdm_bf_update}
\end{algorithm} }

\vspace{-2ex}
\subsection{Complexity Analysis}

A full complexity analysis of the \ac{SotA} scheme of Algorithm 1 is given in \cite{Sandoval23}, such that suffice it for us to consider the overall computational complexity of the \ac{SecLM}-\ac{BF} algorithm proposed here, which can be expressed as
\begin{eqnarray}
\mathcal{O} = \;  i_\text{FP} \cdot i_\text{CCP} \cdot K^2 (N^2 M + N M^2 + N^3) && \nonumber \\ && \hspace{-39.5ex} + i_\text{FP} \cdot i_\text{CCP} \cdot i_\text{BS} \cdot K N^3
\end{eqnarray}
where $i_{\text{FP}}$ is the number of fractional programming iterations, $i_{\text{CCP}}$ is the number of concave-convex procedure iterations, $i_{\text{BS}}$ is the number of bisection search iterations, $K$ is the number of users, $N$ is the total number of transmit antennas, and $M$ is the number of receive antennas per user.

The first term, $i_\text{FP} \cdot i_\text{CCP} \cdot K^2 (N^2 M + N M^2 + N^3)$, accounts for the complexity of computing the rate and leakage terms, their minorization functions, and the gradients for the leakage terms in each iteration, while the second term, $i_\text{FP} \cdot i_\text{CCP} \cdot i_\text{BS} \cdot K N^3$, corresponds to the complexity of the bisection method used to update the beamforming matrices under the power constraint.
%

{
For the \ac{CDM} implementation, the cost difference appears in the bisection-dependent solve of \eqref{eq:bf_update_ccp}.
The auxiliary-variable updates, rate and leakage evaluations, and leakage-gradient computations retain the same order as in the inverse-based \ac{FP}-\ac{CCP} method.
In Algorithm~\ref{alg:cdm_bf_update}, one row-block update has complexity $\mathcal{O}(NM)$ because the residual correction in \eqref{eq:cdm_residual_update} uses one $N$-dimensional column and one $1\times M$ row.
A complete sweep over all $N$ row blocks therefore costs $\mathcal{O}(N^2M)$.
If $T_{\mathrm{CDM}}\leq T_{\max}$ denotes the number of sweeps required to meet the residual tolerance, the linear-solve cost per user and per bisection trial changes from $\mathcal{O}(N^3)$ to $\mathcal{O}(T_{\mathrm{CDM}}N^2M)$.
Accordingly, the complexity of the \ac{CDM}-based \ac{SecLM}-\ac{BF} implementation becomes
\begin{eqnarray}
\label{eq:complexity_cdm}
\mathcal{O}_{\mathrm{CDM}} = i_\text{FP} \cdot i_\text{CCP} \cdot K^2 (N^2M + NM^2) && \nonumber \\ && \hspace{-37ex}
+ i_\text{FP} \cdot i_\text{CCP} \cdot i_\text{BS} \cdot K \cdot T_{\mathrm{CDM}}N^2M.
\end{eqnarray}
The first term in \eqref{eq:complexity_cdm} collects the operations outside the linear solve, whereas the second term is the \ac{CDM} beamformer update over the bisection trials.
Thus, the arithmetic advantage of the \ac{CDM} update is conditional: the bisection-dependent solve is cheaper than the inverse-based update only when $T_{\mathrm{CDM}}M<N$.
If this condition is not met, and in particular if $T_{\mathrm{CDM}}>N$, the asymptotic gain can vanish; the remaining benefits are that \ac{CDM} avoids explicit matrix inversion, uses residual-based stopping, and may improve numerical robustness in ill-conditioned solves.
}

In contrast, the computational complexity of the \ac{SDP}-based \ac{SotA} method is primarily determined by the number of interior-point iterations required to solve the associated SDP problem.
These problems fall into the category of convex determinant maximization, and referring to \cite[Sec.2]{vandenberge98}, the complexity of the \ac{SDP}-based method can be expressed as
\begin{equation}
\mathcal{O} (i_\text{SDP} \cdot K^{4.5} (N^7 M^2)).
\end{equation}
For the sake of completeness, also the complexity of the \ac{SRM}-based \ac{FP} method, which is considered here as \ac{SotA} for sum and secrecy rate maximization in large-scale \ac{CF-mMIMO} systems, is given by
\begin{equation}
O \left( i_{\text{FP}} \cdot i_{\text{BS}} \cdot K^2 (N^2 M + N M^2 + N^3) \right).
\end{equation}

The computational complexity of the proposed \ac{SecLM}-\ac{BF} and the \ac{FP}-based algorithms are compared as functions of the number of transmit antennas.
It is found that the proposed \ac{SecLM}-\ac{BF} exhibits complexity comparable to the \ac{FP}-based approach, significantly lower than the \ac{SDP}-based method, making it scalable for large-scale systems.


\section{Simulation Results}
\label{subsec:result}

In this section, the performance of the proposed \ac{SecLM}-\ac{BF} algorithm is evaluated through simulations.
The simulation parameters are summarized in Table~\ref{table:network-params}.
The simulation is conducted in a square area of $D_{UE} \times D_{UE}$, where the \acp{UE} are randomly distributed, and the \acp{AP} are placed at the corners of a square area of $D_{AP} \times D_{AP}$.
The \acp{AP} are assumed to have a height of $h_{AP} = 10$ m, while the \acp{UE} are at a height of $h_{UE} = 1.5$ m.
The carrier frequency is set to $f_c = 2$ GHz, and the noise variance is set to $\sigma^2 = -96$ dBm.

\begin{figure}[!t]
\centering
\includegraphics[width=1.0\columnwidth]{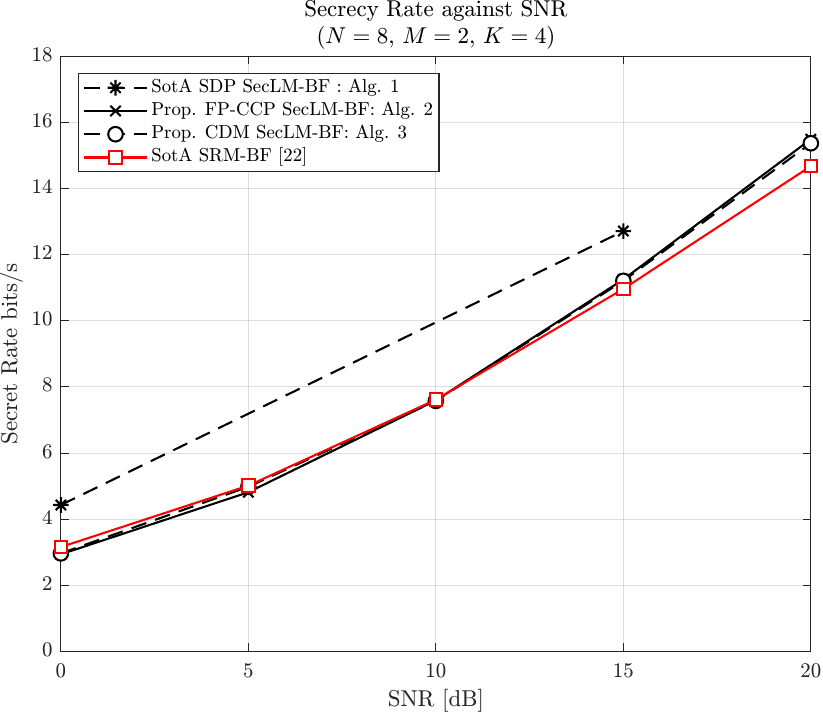}
\vspace{-2ex}
\caption{{Direct \ac{TX}-\ac{BF} secrecy rate versus SNR for $N=8$, $M=2$, and $K=4$}}
\label{fig:secrecy_fullyloaded_BF}
\vspace{0ex}
\includegraphics[width=1.0\columnwidth]{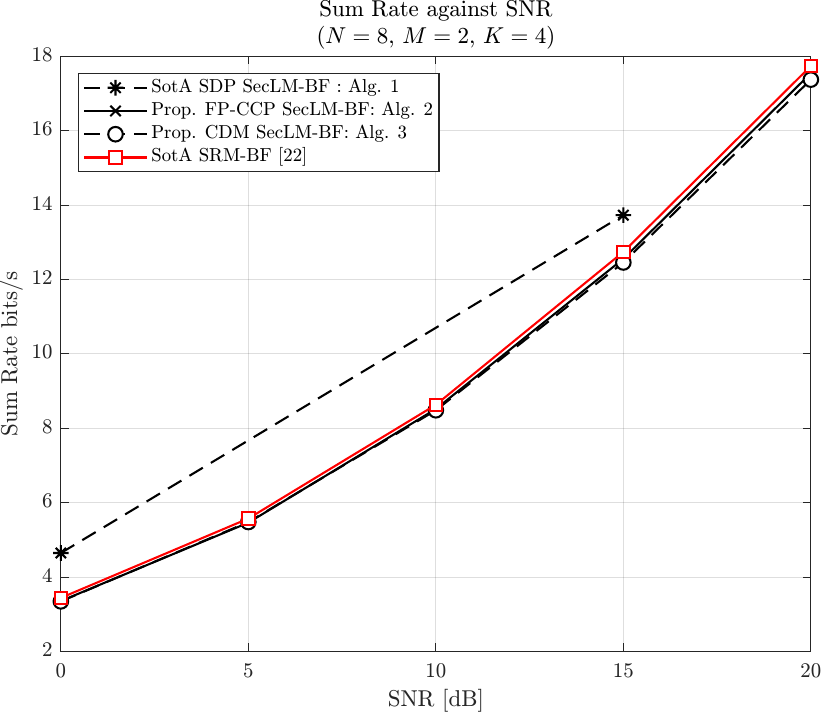}
\vspace{-2ex}
\caption{{Direct \ac{TX}-\ac{BF} communication sum rate versus SNR for $N=8$, $M=2$, and $K=4$}}
\label{fig:sumrate_fullyloaded_BF}
\vspace{-2ex}
\end{figure}

\begin{figure*}[!t]
\centering
\begin{minipage}[t]{0.485\textwidth}
\centering
\subfigure[{\footnotesize{Low SNR underloaded}}]%
{\includegraphics[width=0.96\linewidth]{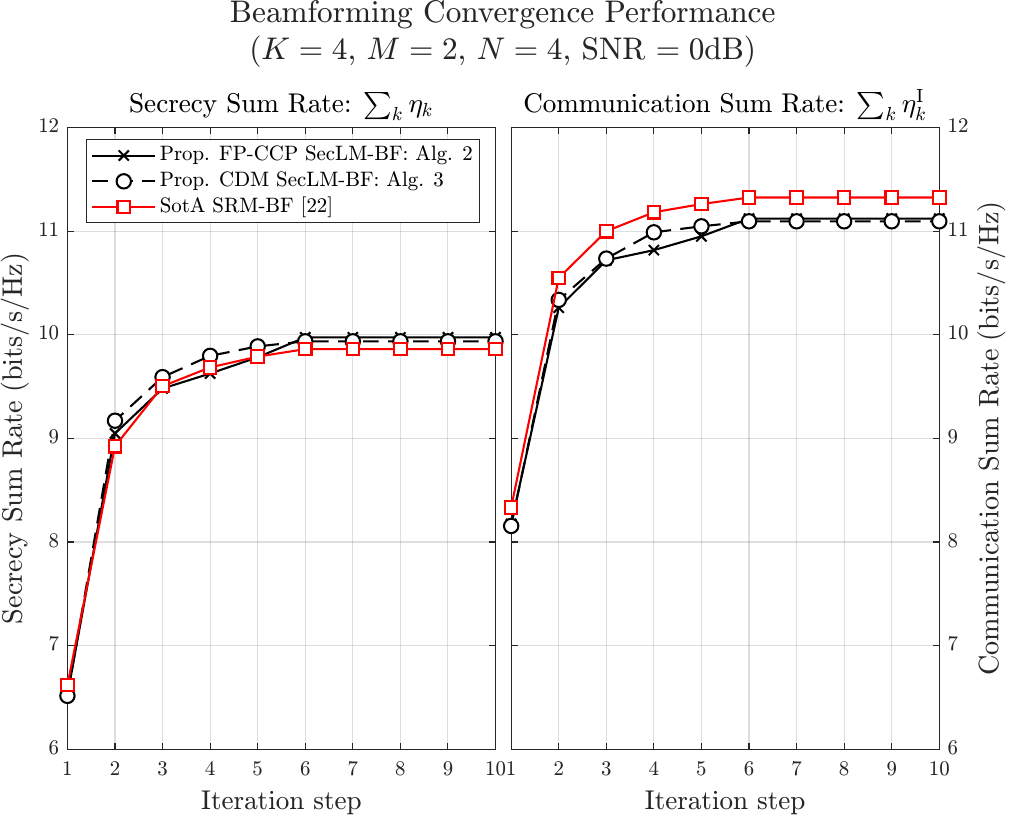}
\label{fig:conv_low_underloaded}}\\[-0.75ex]
\subfigure[{\footnotesize{Low SNR fully loaded}}]%
{\includegraphics[width=0.96\linewidth]{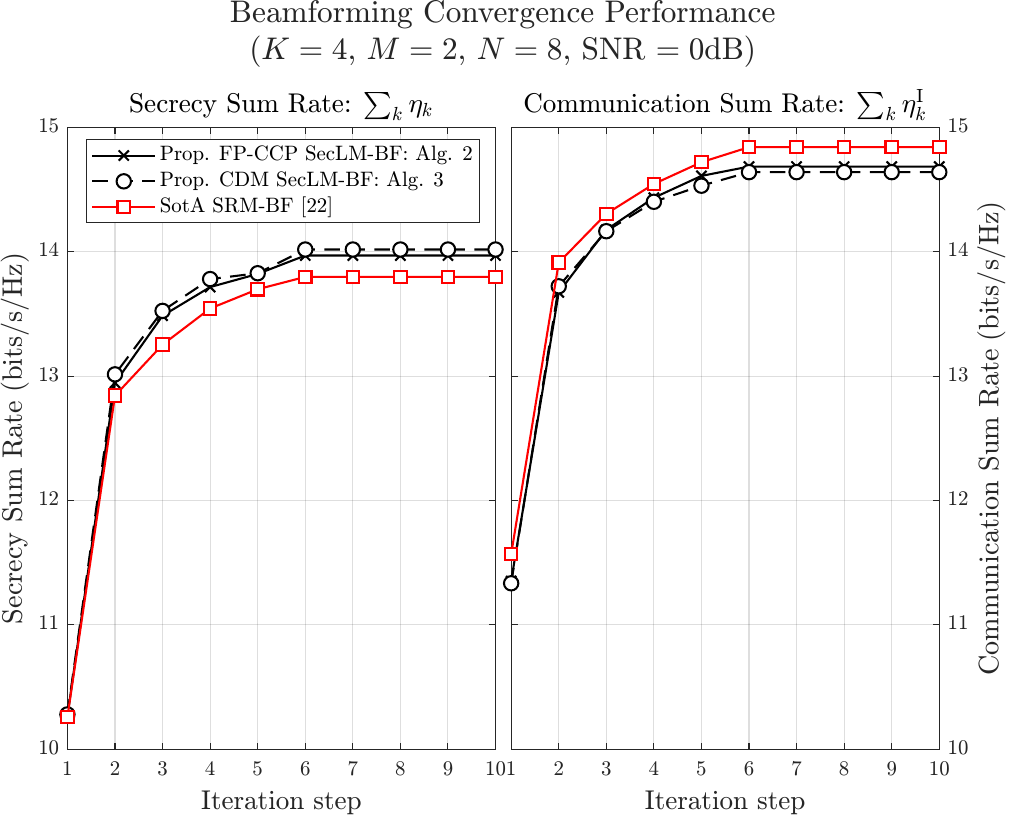}
\label{fig:conv_low_fullyloaded}}\\[-0.75ex]
\subfigure[{\footnotesize{Low SNR overloaded}}]%
{\includegraphics[width=0.96\linewidth]{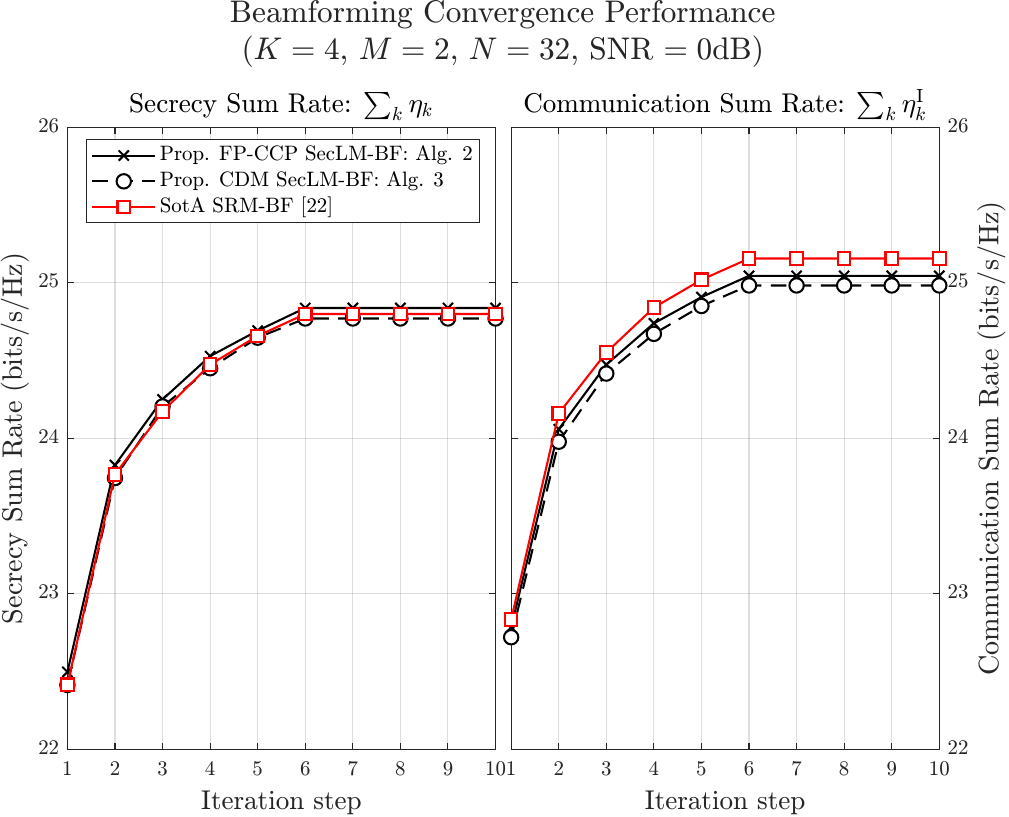}
\label{fig:conv_low_overloaded}}
\vspace{-0.75ex}
\caption{{Low-SNR \ac{BDRIS}/\ac{BF} alternating-optimization behavior across loading regimes}}
\label{fig:convergence_low_snr}
\end{minipage}
\hfill
\begin{minipage}[t]{0.485\textwidth}
\centering
\subfigure[{\footnotesize{High SNR underloaded}}]%
{\includegraphics[width=0.96\linewidth]{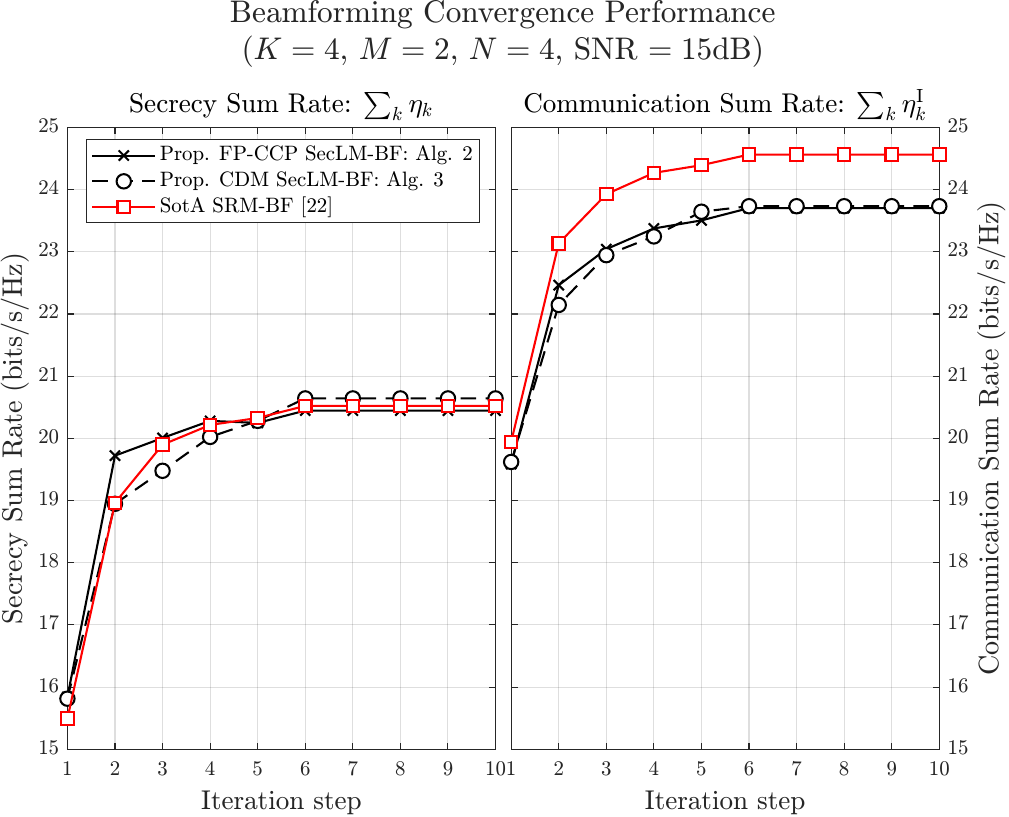}
\label{fig:conv_high_underloaded}}\\[-0.75ex]
\subfigure[{\footnotesize{High SNR fully loaded}}]%
{\includegraphics[width=0.96\linewidth]{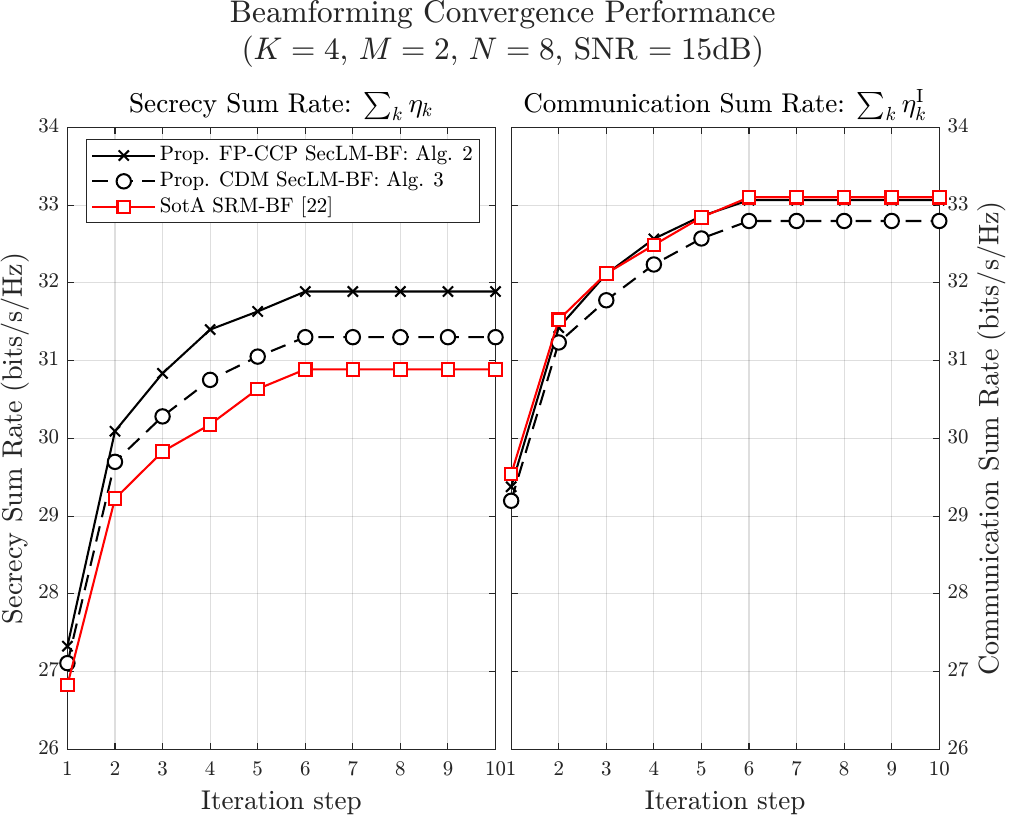}
\label{fig:conv_high_fullyloaded}}\\[-0.75ex]
\subfigure[{\footnotesize{High SNR overloaded}}]%
{\includegraphics[width=0.96\linewidth]{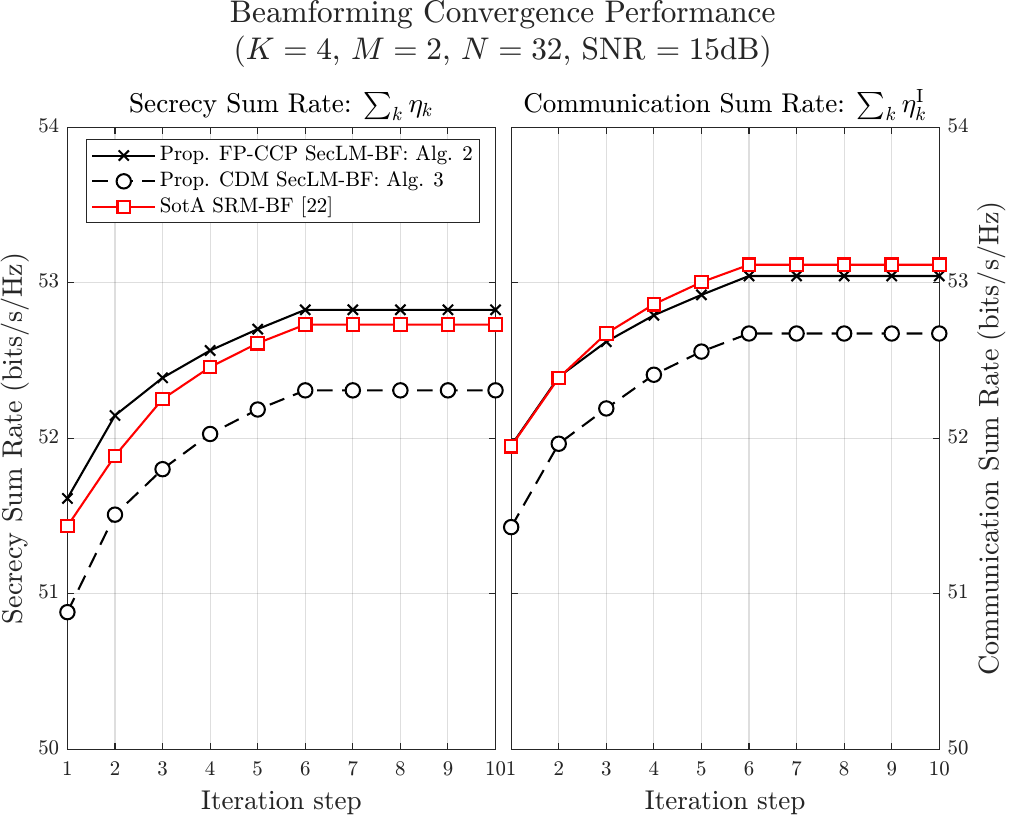}
\label{fig:conv_high_overloaded}}
\vspace{-0.75ex}
\caption{{High-SNR \ac{BDRIS}/\ac{BF} alternating-optimization behavior across loading regimes}}
\label{fig:convergence_high_snr}
\end{minipage}
\vspace*{4ex}
\end{figure*}
\afterpage{%
\begin{figure*}[!t]
\centering
\vspace*{-2ex}
{
\begin{center}
\begin{subequations}
\begin{align}
\mathbf{D}_{\Theta_g}^{\mathrm{I}}
&\triangleq \frac{\ln(2)}{2}\nabla_{\bm{\Theta}_g}\!\sum_{k\in\mathcal{K}}\breve{\eta}_{k}^{\mathrm{I}}
\nonumber\\[-0.5ex]
&=
\sum_{k\in\mathcal{K}}
\mathbf{H}_{\mathrm{RX},k,g}^{\mathrm{H}}\mathbf{Y}_{k}^{\mathrm{I}}
\left(\mathbf{I}_{M}+\mathbf{\Gamma}_{k}^{\mathrm{I}}\right)
\mathbf{V}_{k}^{\mathrm{H}}\mathbf{H}_{\mathrm{TX},g}^{\mathrm{H}}
-\sum_{k\in\mathcal{K}}
\mathbf{H}_{\mathrm{RX},k,g}^{\mathrm{H}}\mathbf{Y}_{k}^{\mathrm{I}}
\left(\mathbf{I}_{M}+\mathbf{\Gamma}_{k}^{\mathrm{I}}\right)
{\mathbf{Y}_{k}^{\mathrm{I}}}^{\mathrm{H}}
\bHeq{k}\mathbf{S}\mathbf{H}_{\mathrm{TX},g}^{\mathrm{H}}. \label{eq:bdris_secrecy_theta_gradient_intended}
\end{align}
\rule{0.5\textwidth}{0.1pt}
\begin{align}
\mathbf{D}_{\Theta_g}^{\mathrm{L}}
&\triangleq \frac{\ln(2)}{2}\nabla_{\bm{\Theta}_g}\!\sum_{k\in\mathcal{K}}\breve{\eta}_{k,\tilde e_k}^{\mathrm{L}}
\nonumber\\[-0.5ex]
&=
\sum_{k\in\mathcal{K}}
\mathbf{H}_{\mathrm{RX},\tilde e_k,g}^{\mathrm{H}}\mathbf{Y}_{k,\tilde e_k}^{\mathrm{L}}
\left(\mathbf{I}_{M}+\mathbf{\Gamma}_{k,\tilde e_k}^{\mathrm{L}}\right)
\mathbf{V}_{k}^{\mathrm{H}}\mathbf{H}_{\mathrm{TX},g}^{\mathrm{H}}
-\sum_{k\in\mathcal{K}}
\mathbf{H}_{\mathrm{RX},\tilde e_k,g}^{\mathrm{H}}\mathbf{Y}_{k,\tilde e_k}^{\mathrm{L}}
\left(\mathbf{I}_{M}+\mathbf{\Gamma}_{k,\tilde e_k}^{\mathrm{L}}\right)
{\mathbf{Y}_{k,\tilde e_k}^{\mathrm{L}}}^{\mathrm{H}}
\bHeq{\tilde e_k}\mathbf{S}_{\tilde e_k}^{(-)}\mathbf{H}_{\mathrm{TX},g}^{\mathrm{H}}. \label{eq:bdris_secrecy_theta_gradient_leakage}
\end{align}
\end{subequations}
\end{center}
}
\vspace{0.25ex}
\rule{1\textwidth}{0.1pt}
\par\vspace{-5.25ex}\noindent\mbox{}
\end{figure*}
}
{
The numerical results in Figs.~\ref{fig:secrecy_fullyloaded_BF}--\ref{fig:convergence_high_snr} evaluate the proposed \ac{SecLM}-\ac{BF} design under two channel settings.
Figs.~\ref{fig:secrecy_fullyloaded_BF} and~\ref{fig:sumrate_fullyloaded_BF} consider direct \ac{TX} \ac{BF} without a \ac{BDRIS}, thereby isolating the beamformer behavior over the direct \ac{AP}-\ac{UE} channels, whereas Figs.~\ref{fig:convergence_low_snr} and~\ref{fig:convergence_high_snr} consider the \ac{BDRIS}-assisted setting.
In Figs.~\ref{fig:convergence_low_snr} and~\ref{fig:convergence_high_snr}, the first plotted point corresponds to the initial beamformer before the first \ac{BDRIS} update, and each subsequent point corresponds to one completed \ac{BDRIS}/\ac{BF} alternating-optimization cycle.
This plotted iteration index is distinct from the inner \ac{CDM} sweep count $T_{\mathrm{CDM}}$ and from the inner \ac{BDRIS} phase-update iterations.
}

{
For the \ac{BDRIS}-assisted evaluation, the antenna loading is varied across the overloaded, fully loaded, and underloaded regimes by setting $N=4$, $N=8$, and $N=32$, respectively.
The \ac{BDRIS} is modeled as a single-connected surface with $N_{\mathrm{RIS}}=32$, $G=32$, $n_g=1$, and height $h_{\mathrm{RIS}}=5$ m\footnote{{This scenario is considered due to its low complexity; it should be noted, however, that the considered framework is general, and allows for the usage of more complex \ac{RIS} configurations, such as the group- and fully-connected.}}.
For all \ac{CDM} beamformer updates, the maximum number of row sweeps is set to $T_{\max}=50$, with residual tolerance $\epsilon_{\mathrm{CDM}}=10^{-8}$ and warm starting enabled across successive bisection trials.
In a representative fully loaded \ac{BDRIS}-assisted run at $0$~dB, the residual-converged \ac{CDM} solves typically required only a few sweeps, with a median of three sweeps, although difficult bisection trials may reach the $50$-sweep cap.
Because the \ac{BDRIS} phase-update convergence has been examined in prior work \cite{SandovalArx2026}, this section focuses on the resulting secrecy and sum-rate trends rather than on the inner \ac{BDRIS} iteration budget.
This separation allows Figs.~\ref{fig:secrecy_fullyloaded_BF} and~\ref{fig:sumrate_fullyloaded_BF} to assess the \ac{TX} beamformer alone, while Figs.~\ref{fig:convergence_low_snr} and~\ref{fig:convergence_high_snr} show the corresponding behavior in the presence of the \ac{BDRIS}.

The \ac{BDRIS} scattering-matrix update adopts the reciprocal \ac{BDRIS} design of~\cite{SandovalArx2026}; only the Euclidean gradient is replaced by the gradient of the \ac{FP} secrecy surrogate in \eqref{eq:secbf_FP_obj}, which follows from the secrecy-rate objective in \eqref{eq:priv_secbf_sdp_prop_obj} after the auxiliary-variable transformations, while the group-wise manifold projection, Armijo search, and retraction remain unchanged.
For convenience, we denote the concatenated channel from all $L$ \acp{AP} to the \ac{BDRIS} as $\mathbf{H}_{\mathrm{TX}}$, where 
\begin{equation}
\mathbf{H}_{\mathrm{TX}} \triangleq [\bHTX{1},\ldots,\bHTX{L}].
\end{equation}

For the group-connected architecture, the \ac{BDRIS} scattering matrix is optimized through the diagonal scattering blocks associated with the disjoint element groups, namely
\begin{equation}
\begin{aligned}
\bTheta&=\operatorname{blkdiag}\left(\bm{\Theta}_1,\ldots,\bm{\Theta}_G\right),
\\
\bm{\Theta}_g&\in\mathrm{St}(n_g,n_g;\mathbb{C}),\quad
g=1,\ldots,G.
\end{aligned}
\end{equation}

Let $\mathcal{R}_g=\{(g-1)n_g+1,\ldots,gn_g\}$ denote the \ac{BDRIS} element indices of group $g$.
We define the corresponding channel slices as $\mathbf{H}_{\mathrm{RX},k,g}\triangleq[\bHRX{k}]_{:,\mathcal{R}_g}$ and $\mathbf{H}_{\mathrm{TX},g}\triangleq[\mathbf{H}_{\mathrm{TX}}]_{\mathcal{R}_g,:}$.
We also define the total transmit covariance $\mathbf{S} \triangleq \sum_{i\in\mathcal{K}}\mathbf{V}_i\mathbf{V}_i^{\mathrm{H}}$ and the leakage covariance excluding the $\tilde e_k$-th user as $\mathbf{S}_{\tilde e_k}^{(-)} \triangleq \mathbf{S}-\mathbf{V}_{\tilde e_k}\mathbf{V}_{\tilde e_k}^{\mathrm{H}}$.
Accordingly, the Euclidean gradient with respect to the $g$-th scattering block can be expressed as
\begin{equation}
\label{eq:bdris_secrecy_theta_gradient_main}
\begin{aligned}
\nabla_{\bm{\Theta}_g}\breve{\eta}^{\mathrm{sec}}
=
\frac{2}{\ln(2)}\big(\mathbf{D}_{\Theta_g}^{\mathrm{I}}\!\!-\!\mathbf{D}_{\Theta_g}^{\mathrm{L}}\big) \!-4\alpha_{\mathrm{sym}}\big(\bm{\Theta}_g\!-\!\bm{\Theta}_g^{\mathrm{T}}\big),
\end{aligned}
\end{equation}
for all $g=1,\ldots,G$, and where $\mathbf{D}_{\Theta_g}^{\mathrm{I}}$ and $\mathbf{D}_{\Theta_g}^{\mathrm{L}}$ are the scaled intended-rate and leakage-rate gradient components defined in \eqref{eq:bdris_secrecy_theta_gradient_intended} and \eqref{eq:bdris_secrecy_theta_gradient_leakage}, seen on the next page of the article, respectively.
The derivative in \eqref{eq:bdris_secrecy_theta_gradient_main} is therefore taken with respect to the group variable $\bm{\Theta}_g$ itself, while entries outside the diagonal group blocks are excluded by the group-connected feasible set.
The Riemannian gradient used by the \ac{CGA} step is then obtained by projecting each Euclidean block gradient onto the tangent space of $\mathrm{St}(n_g,n_g;\mathbb{C})$ as
\begin{equation}
\label{eq:bdris_group_riemannian_gradient}
\mathrm{T}_{\bm{\Theta}_g}
=
\nabla_{\bm{\Theta}_g}\breve{\eta}^{\mathrm{sec}}-\bm{\Theta}_g
\frac{\bm{\Theta}_g^{\mathrm{H}}\nabla_{\bm{\Theta}_g}\breve{\eta}^{\mathrm{sec}}+\nabla_{\bm{\Theta}_g}\breve{\eta}^{\mathrm{sec}, \mathrm{H}}\bm{\Theta}_g}{2}.
\end{equation}

The Armijo search and QR-based retraction are applied independently to the blocks $\bm{\Theta}_g$, after which $\bTheta$ is assembled again as a block-diagonal matrix.
The single-connected, group-connected, and fully connected cases are recovered by setting $n_g=1$, using an intermediate block size, or setting $G=1$ and $n_g=N_{\mathrm{RIS}}$, respectively.

In Fig. \ref{fig:secrecy_fullyloaded_BF} and \ref{fig:sumrate_fullyloaded_BF}, the proposed \ac{SecLM}-\ac{BF} algorithms are benchmarked against two \ac{SotA} methods, $i.e.$, the \ac{SDP}-based leakage-minimization method in~\cite{Sandoval23}, and the \ac{FP}-based sum-rate maximization method based on~\cite{shen2018-II} and developed in~\cite{Sandoval23}.
The number of \ac{CCP} iterations is set to $I_{\text{CCP}}=10$, with convergence tolerances $\epsilon_{\text{CCP}}=10^{-3}$ and $\epsilon_{\text{FP}}=10^{-3}$.

The \ac{SDP}-based benchmark gives the largest secrecy rate at the sampled \ac{SNR} points, while the proposed curves remain close without requiring an interior-point solution.
The small high-\ac{SNR} separation from \ac{SRM}-\ac{BF} reflects the secrecy-throughput tradeoff, since \ac{SecLM} also suppresses leakage whereas \ac{SRM} mainly increases useful-link throughput.

Furthermore, the direct-channel curves show that the \ac{CDM} implementation closely follows the inverse-based \ac{FP}-\ac{CCP} update in both secrecy rate and communication sum rate across the sampled \ac{SNR} range.
This agreement indicates that the residual-controlled row sweeps preserve the beamformer quality of the inverse-based update before the additional \ac{BDRIS}-assisted propagation path is introduced.
}

{
The \ac{BDRIS}-assisted alternating-optimization results in Fig.~\ref{fig:convergence_low_snr} and~\ref{fig:convergence_high_snr} compare the low-complexity \ac{FP}-\ac{CCP}, \ac{CDM}, and \ac{SRM}-\ac{BF} methods under the same reflected-channel conditions.
The \ac{SDP} curve is not included in these plots due to the excessively high computational cost of the \ac{SDP} update, which would require a separate run for each point in the alternating-optimization trace.
These two groups of figures therefore highlight complementary effects rather than repeating the same comparison.
The direct \ac{SNR} sweep in Fig.~\ref{fig:secrecy_fullyloaded_BF} and~\ref{fig:sumrate_fullyloaded_BF} checks whether the inversion-free \ac{CDM} update changes the rate behavior when only the direct \ac{AP}-\ac{UE} channels are active.
}
\begin{table}[t]
\small
\centering
\caption{Simulation Parameters}
\vspace{-1ex}
\label{table:network-params}
\renewcommand{\arraystretch}{1.2}
\begin{tabularx}{0.81\columnwidth}{clccc}
\hline
\multicolumn{1}{c}{Symbol} & \multicolumn{1}{c}{Meaning} & \multicolumn{3}{c}{Values} \\ \hline
$L$                 & Number of \acp{AP} & \multicolumn{3}{c}{4}\\
$K$                 & Number of \acp{UE} & \multicolumn{3}{c}{4}\\
$M$                 & \ac{UE} antennas   & \multicolumn{3}{c}{2}\\
$N$                 & Total \ac{AP} antennas & {4} & {8} & {32}\\
{$N_{\mathrm{RIS}}$} & {BD-RIS elements} & \multicolumn{3}{c}{{32}}\\
{$G,n_g$} & {BD-RIS grouping} & \multicolumn{3}{c}{{32, 1}}\\
$h_{AP}$            & \ac{AP} height & \multicolumn{3}{l}{10[m]}\\
$h_{UE}$            & \ac{UE} height & \multicolumn{3}{l}{1.5[m]}\\
{$h_{\mathrm{RIS}}$} & {BD-RIS height} & \multicolumn{3}{l}{{5[m]}}\\
$f_c$               & Carrier frequency & \multicolumn{3}{l}{2[GHz]}\\
$D_{UE}$            & UE square side length & \multicolumn{3}{l}{500[m]}\\
$D_{AP}$            & AP square side length & \multicolumn{3}{l}{300[m]}\\
$\sigma^2$          & Receiver noise variance & \multicolumn{3}{l}{-96[dBm]} \\
{$T_{\max}$} & {CDM sweep cap} & \multicolumn{3}{l}{{$50$}}\\
{$\epsilon_{\mathrm{CDM}}$} & {CDM residual tolerance} & \multicolumn{3}{l}{{$10^{-8}$}}\\
\hline
\end{tabularx}
\vspace{-3ex}
\end{table}

{
\vspace{-2ex}
At $0$~dB, Fig.~\ref{fig:convergence_low_snr} shows that all \ac{BDRIS}-assisted low-complexity methods reach stable values within the first six plotted points, corresponding to the initialization and the first five \ac{BDRIS}/\ac{BF} alternating-optimization cycles.
In the overloaded case in Fig.~\ref{fig:conv_low_overloaded}, the proposed \ac{FP}-\ac{CCP} and \ac{CDM} methods achieve a slightly higher secrecy sum rate than the \ac{SRM}-\ac{BF} benchmark, while maintaining a comparable communication sum rate.
The same trend is more evident in the fully loaded case in Fig.~\ref{fig:conv_low_fullyloaded}, where the proposed secrecy-aware methods converge above \ac{SRM}-\ac{BF} in secrecy sum rate and only slightly below it in communication sum rate.
In the underloaded case in Fig.~\ref{fig:conv_low_underloaded}, the performance differences become small because the larger number of transmit antennas and the \ac{BDRIS}-assisted reflected path provide sufficient spatial degrees of freedom for effective user separation.
Thus, the proposed method provides the clearest secrecy gain when spatial resources are limited, whereas in the underloaded regime its main advantage is to preserve high performance with a low-complexity \ac{FP}-based implementation.

At $15$~dB, Fig.~\ref{fig:convergence_high_snr} shows that the secrecy benefit of the proposed objective becomes more pronounced in the \ac{BDRIS}-assisted setting.
In the overloaded case in Fig.~\ref{fig:conv_high_overloaded}, the proposed methods converge to a secrecy sum rate slightly above the \ac{SRM}-\ac{BF} benchmark, although \ac{SRM}-\ac{BF} achieves a higher communication sum rate.
This behavior reflects the expected tradeoff, since the \ac{SRM} objective uses the additional \ac{SNR} mainly to increase useful-link throughput, whereas the proposed \ac{SecLM} objective also suppresses the strongest leakage paths.
In the fully loaded case in Fig.~\ref{fig:conv_high_fullyloaded}, the proposed secrecy-based beamformer achieves a communication sum rate that is almost identical to that of \ac{SRM}-\ac{BF}, while providing a clear secrecy sum-rate gain.
This result shows that, when the system is fully loaded, the proposed objective can suppress leakage without sacrificing the useful-link throughput targeted by the \ac{SRM} design.
In the underloaded case in Fig.~\ref{fig:conv_high_underloaded}, the inverse-based \ac{FP}-\ac{CCP} and \ac{SRM}-\ac{BF} curves remain close because the additional antenna and \ac{BDRIS} degrees of freedom already provide strong user separation.
In this regime, the benefit of the proposed secrecy-aware design is therefore reflected mainly in preserving the near-\ac{SRM} communication sum rate while maintaining explicit leakage control, whereas the \ac{CDM} curve illustrates the performance cost of the inversion-free row-wise update.

Overall, Fig.~\ref{fig:secrecy_fullyloaded_BF}--\ref{fig:convergence_high_snr} show that the proposed \ac{SecLM}-\ac{BF} framework provides a favorable balance between secrecy performance, communication throughput, and computational scalability.
The \ac{CDM} implementation is attractive because it closely matches the inverse-based \ac{FP}-\ac{CCP} behavior in the direct fully loaded \ac{SNR} sweep and provides comparable \ac{BDRIS}-assisted alternating-optimization trends across the antenna-loading regimes, while replacing direct matrix inversions with coordinate-wise linear-system updates.
Hence, when the secrecy gains are large, they are mainly due to leakage-aware beam shaping, and when the curves are similar, the proposed method still offers an inversion-free, residual-controlled update that can reduce arithmetic cost when the \ac{CDM} sweep count remains moderate.
}

\vspace{-2ex}
\section{Conclusion}
\label{sec:conclusion}

In this paper, we proposed a novel beamforming design for secrecy rate maximization in large-scale \ac{CF-mMIMO} systems, termed \ac{SecLM}-\ac{BF}.
The proposed method leverages a combination of \ac{FP}, \ac{CCP}, {and \ac{CDM}} to efficiently solve the non-convex optimization problem associated with secrecy rate maximization.
The algorithm iteratively refines the beamforming matrices while ensuring that the transmit power constraint is satisfied and the secrecy rate is maximized.
{
The simulation results demonstrate that the proposed \ac{SecLM}-\ac{BF} framework achieves competitive secrecy rates and communication sum rates compared with \ac{SotA} benchmarks, while the \ac{CDM} implementation closely tracks the inverse-based update in the tested regimes and avoids explicit matrix inversion.
%
\vspace{-2ex}
}

\bibliographystyle{template/IEEEtran}
\bibliography{listofpublications}

\vspace{-5ex}
\begin{IEEEbiography}[{\includegraphics[width=1in,height=1.25in,trim=19.2bp 0 19.2bp 0,clip]{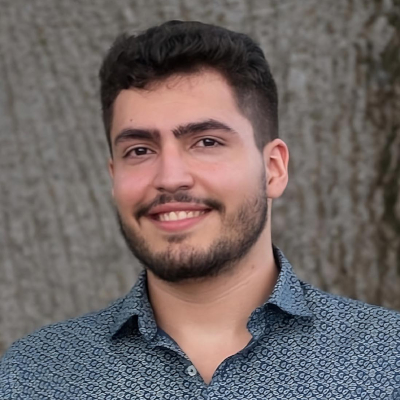}}]{Iv{\'a}n Alexander Morales Sandoval}
(Graduate Student Member, IEEE) received the B.Sc. degree in Electrical and Computer Engineering from Constructor University, Bremen, Germany, in 2018, where he is currently pursuing a Ph.D. degree in electrical engineering. His research interests include signal processing, wireless physical layer key generation for resource constrained devices and wireless physical layer secrecy maximization and authentication schemes.
\end{IEEEbiography}
\vspace{-6ex}

\begin{IEEEbiography}[{\includegraphics[width=1in,height=1.25in,trim=0 37.5bp 0 37.5bp,clip]{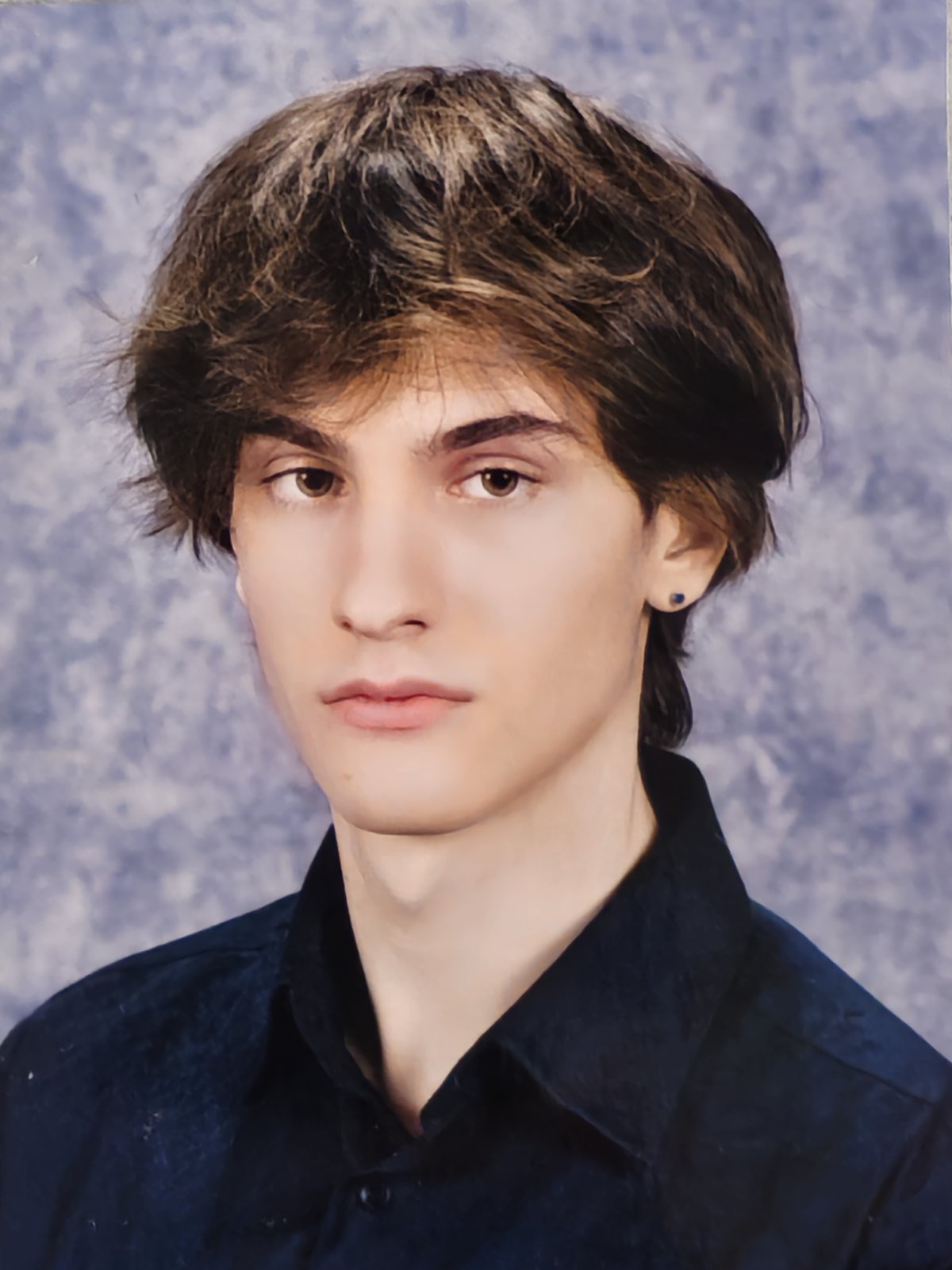}}]{Marko Fidanovski}
(Graduate Student Member, IEEE) received the B.Sc. degree in Electrical and Computer Engineering from Constructor University (formerly Jacobs University Bremen), Bremen, Germany, in 2025. He is currently pursuing the M.Sc. degree in Communications and Electronics Engineering at the Technical University of Munich, Munich, Germany. Since 2025, he has been a Research Associate with the School of Computer Science and Engineering, Constructor University. His research interests include wireless communications and signal processing, with a focus on reciprocal beyond-diagonal reconfigurable intelligent surfaces, manifold optimization, fractional programming, MIMO beamforming, cell-free massive MIMO, and physical-layer security.
\end{IEEEbiography}
\vspace{-6ex}

\begin{IEEEbiography}[{\includegraphics[width=1in,height=1.25in,trim=20.7bp 0 20.7bp 0,clip]{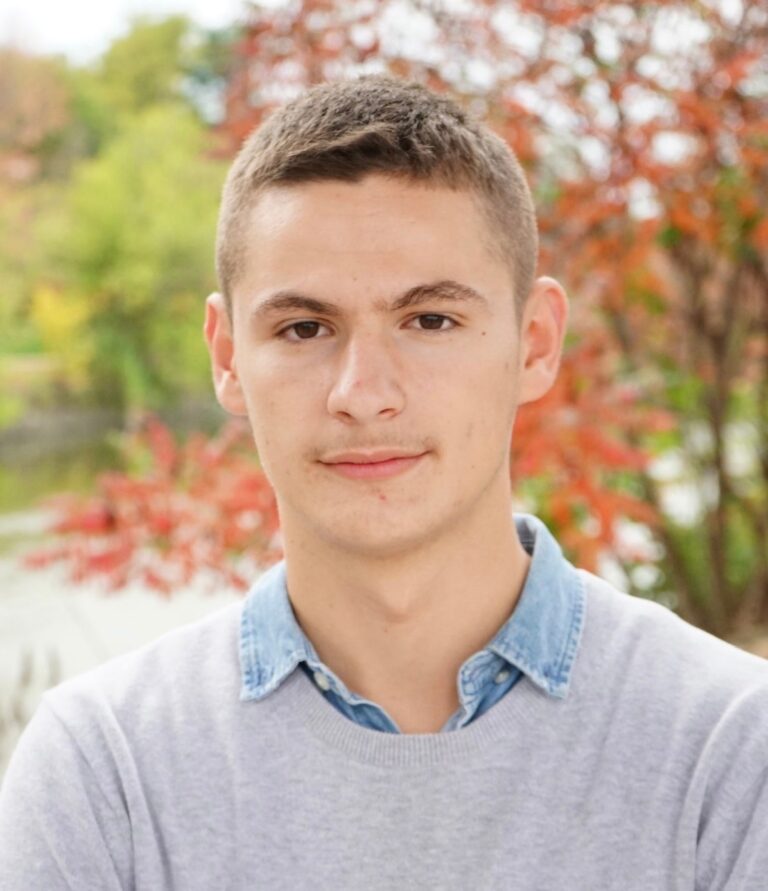}}]{Getuar Rexhepi}
(Graduate Student Member, IEEE) received the B.Sc. degree in Electrical and Computer Engineering from Constructor University (formerly Jacobs University Bremen), Germany, in 2025, where he is currently pursuing the Ph.D. degree in Electrical Engineering. He has been a visiting researcher at ETH Z{\"u}rich, Switzerland, and KTH Royal Institute of Technology, Sweden. He is currently working with Lenovo on sensing capabilities for IEEE 802.11bf. His research interests include wireless communications and signal processing, with a focus on integrated sensing and communications (ISAC), OFDM waveform design, PAPR reduction, manifold optimization, and next-generation wireless technologies.
\end{IEEEbiography}
\vspace{-6ex}

\begin{IEEEbiography}[{\includegraphics[width=1in,height=1.25in,trim=0 17.125bp 0 17.125bp,clip]{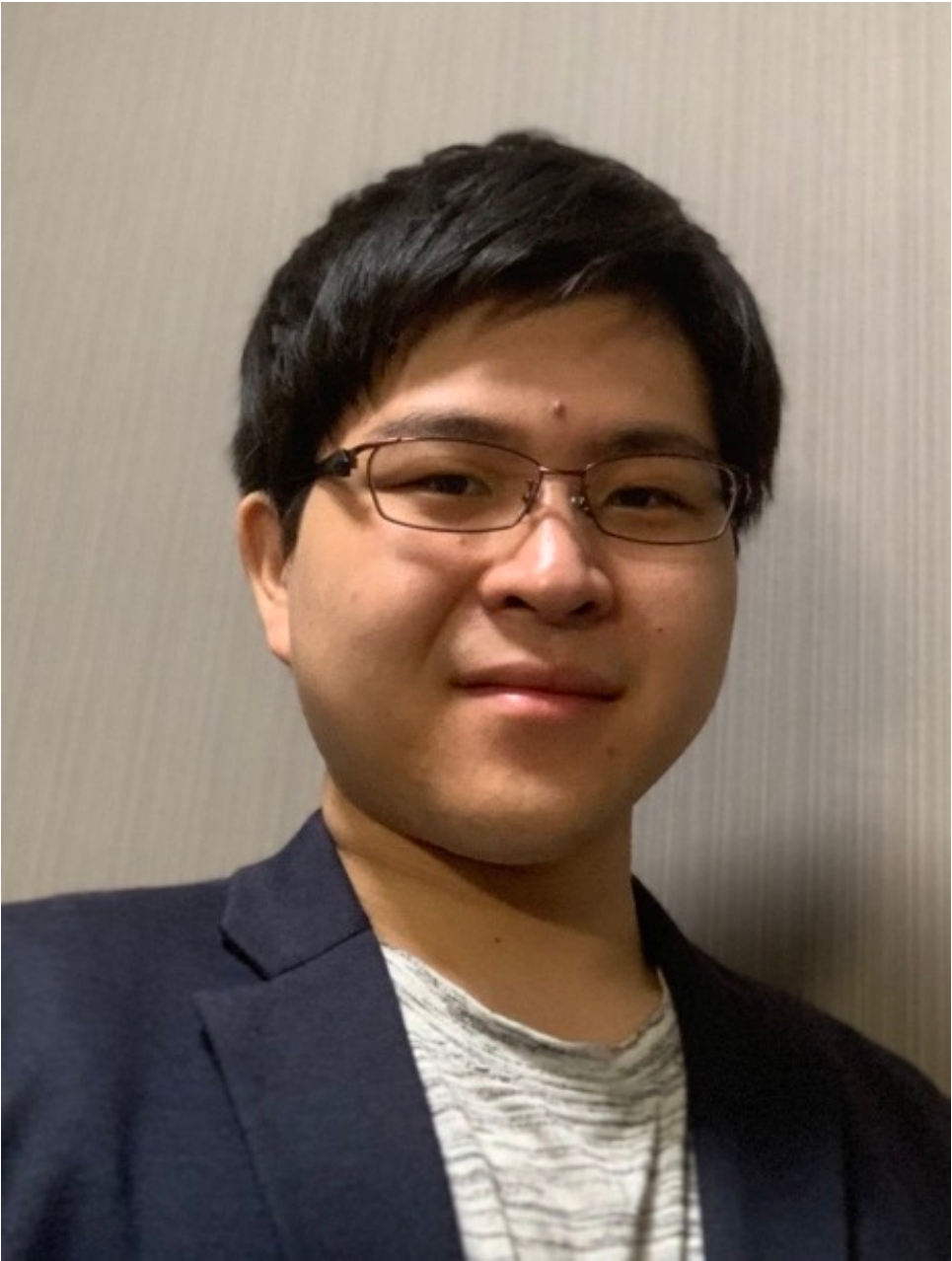}}]{Kengo Ando}
(Member, IEEE) received the B.E. and M.E. degrees in engineering from The University of Electro-Communications, Tokyo, Japan, in 2020 and 2022, respectively. He is currently pursuing the Ph.D. degree in electrical and computer engineering with Constructor University (previously Jacobs University Bremen), Germany. His current research interests are wireless communications and signal processing. He was a recipient of the YKK Graduate Fellowship for Master's students (2020--2022) from the Yoshida Scholarship Foundation, Japan, and the fellowship for overseas study in 2022 from the KDDI Foundation, Japan.
\end{IEEEbiography}


\begin{IEEEbiography}[{\includegraphics[width=1in,height=1.25in,trim=9.6bp 0 9.6bp 0,clip]{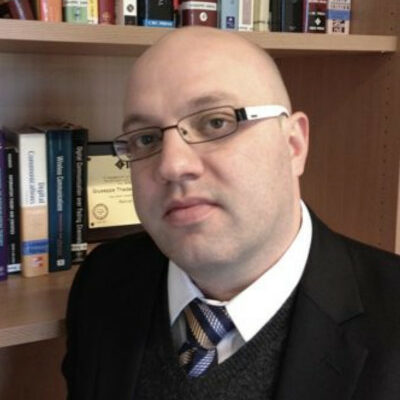}}]{Giuseppe Thadeu Freitas de Abreu}(Senior Member, IEEE) received the B.Eng. degree in electrical engineering and the specialization Latu Sensu
degree in telecommunications engineering from the
Universidade Federal da Bahia (UFBa), Salvador,
Bahia, Brazil in 1996 and 1997, respectively, and
the M.Eng. and D.Eng. degrees in physics, electrical,
and computer engineering from Yokohama National
University, Japan, in March 2001 and March 2004,
respectively.
He was a postdoctoral fellow and later
an adjunct professor (docent) in statistical signal
processing and communications theory at the Department of Electrical and
Information Engineering, University of Oulu, Finland from 2004 to 2006
and from 2006 to 2011, respectively.
Since 2011, he has been a professor
of electrical engineering at Constructor University Bremen (formerly Jacobs University, Bremen, Germany). From April 2015 to August 2018, he simultaneously held a full tenured professorship also
at the Department of Computer and Electrical Engineering, Ritsumeikan
University, Japan.
He has been a visiting scholar at Rice University (2008), Stanford University (2009), Keio University (2010), the University of Bologna (2012 and 2014) and at the KTH Royal Institute of Technology (2025). He has also been a regular visiting professor at the University of Electro-communications, Japan since 2013.
He received the Uenohara Award at Tokyo University in 2000 for his master’s thesis work and was the co-recipient of awards from various international IEEE conferences, including the Asilomar in 2009, 2012 and 2014, WPNC in 2012 and 2015, ISWCS in 2016, WPMC in 2023 and ICNC in 2025.
He was also a co-recipient of the Best Journal Paper Award by the Japanese Chapter of the IEEE Signal Processing Society in 2023 and was awarded various prestigious Fellowships from the Heiwa Nakajima Foundation, the JSPS, and the NICT (twice), respectively in 2010, 2013, 2015 and 2018.
He served as an editor for the IEEE Transactions on Wireless Communications from 2009 to 2014, editor for the IEEE Transactions on Communications from 2014 to 2017, executive editor 
for the IEEE Transactions on Wireless Communications from 2017 to 2021, editor for the IEEE Communication Letters from 2021 to 2024, and is currently serving as an editor to the IEEE Signal Processing Letters and the IEEE Open Journal of the Communications Society.
His research interests include a wide range of topics on wireless communications and signal processing, including communications theory, estimation theory, optimization theory, statistical modeling, metasurfaces and parameterizable electromagnetic structures, waveform design, integrated sensing and communications, over-the-air computing, wireless localization, cognitive radio, wireless security, MIMO systems, ultrawideband and millimeter wave communications, full-duplex and cognitive radio, compressive sensing, energy harvesting networks, random networks, connected vehicles networks, and many more.
\end{IEEEbiography}

\EOD

\end{document}